\documentclass[12pt]{elsarticle}
\usepackage{nomencl}
\usepackage{ifthen}
\renewcommand{\nomgroup}[1]{%
\ifthenelse{\equal{#1}{S}}{\item[\textbf{Subscripts}]}{%
\ifthenelse{\equal{#1}{V}}{\item[\textbf{Variables}]}{%
\ifthenelse{\equal{#1}{S}}{\item[\textbf{sets}]}{}}}
}
\makenomenclature

\usepackage{amssymb}
\usepackage[a4paper]{geometry}  % imports CLiC-it 2023 layout style
\usepackage{times} % font 
\usepackage{xurl} % splits URL in multiple lines
\usepackage[italian,english]{babel}
\usepackage{latexsym} 
\usepackage{xcolor}

\makeatletter
\newcommand{\@BIBLABEL}{\@emptybiblabel}
\newcommand{\@emptybiblabel}[1]{}
\usepackage{graphicx}
\usepackage{latexsym}
\usepackage{amssymb}
\usepackage{bm}
\usepackage{amsmath}
\usepackage{epsfig}
\usepackage{caption}
\usepackage{subcaption}
\usepackage{graphicx}
\usepackage{caption}
\usepackage{subcaption}
\usepackage{mwe} 
\usepackage{morefloats}
\usepackage{xcolor}
\usepackage{epstopdf}
\begin{document}
\begin{frontmatter}


\title{New Higher-Order Super-Compact Scheme for Enhanced Three-Dimensional Heat Transfer with Nanofluid and Conducting Fins}

\makeatletter
\def\ps@pprintTitle{%
  \let\@oddhead\@empty
  \let\@evenhead\@empty
  \let\@oddfoot\@empty
  \let\@evenfoot\@oddfoot
}
% Change the footnote mark to an asterisk
\renewcommand{\thefootnote}{*}
\makeatother
\author{\textbf{Ashwani Punia$^{1}$, Rajendra K. Ray$^{2}$}\footnote{Corresponding author : Rajendra K. Ray, rajendra@iitmandi.ac.in} \\
  1,2. School of Mathematical and Statistical Sciences, Indian Institute of Technology Mandi,\\
  Mandi (H.P), 175075, India \\
 {\tt mr.punia11@gmail.com}, {\tt rajendra@iitmandi.ac.in}}

\begin{abstract}
This study presents a new higher-order super-compact (HOSC) finite difference scheme for analyzing enhanced heat transfer of three-dimensional (3D) nanofluid natural convection in a cubic cavity. The unique contribution of the present work lies in the extension of the higher-order super-compact finite difference scheme to examine the natural convection of nanofluid in the 3D cavity. This numerical approach achieves fourth-order spatial accuracy and second-order temporal accuracy. `Super-compact' term signifies its efficiency, utilizing 19 grid points at the current time level $(n^{th}$ time level$)$ and just seven grid points at the subsequent time level $((n + 1)^{th}$ time level$)$ around which the finite difference discretization is made. The nanoparticle volume fraction is maintained up to 0.04 (4\%) to ensure the mixture exhibits Newtonian behavior. The newly developed numerical scheme is validated by qualitative and quantitative comparisons with existing benchmark results. The scheme is then applied to investigate fluid flow and heat transfer phenomena in a Cu-water nanofluid-filled cavity over a range of Rayleigh numbers ($10^2 \leq Ra \leq 10^5$).  In addition to introducing the new HOSC scheme for the convection of nanofluids, we examine two cases: the natural convection of nanofluid in a simple 3D cavity, and a configuration incorporating two aluminum conducting fins on the heated wall to further enhance the heat transfer rate. Results are presented through isotherms, streamlines, local Nusselt numbers, and average Nusselt numbers for both the considered cases and compared their results. Present computed results reveal significant improvements in thermal performance due to the dual fin configuration. Incorporating the two conducting fins can enhance the heat transfer rate by up to 88.9\% at $Ra=10^2$. It is found that the addition of nanoparticles or conducting fins does not always lead to enhanced heat transfer rates. Instead, the effectiveness of these enhancements is highly dependent on a range of parameters, which are thoroughly examined and discussed in this work. 
\end{abstract}

\end{frontmatter}

\nomenclature{3D}{Three-dimensional}
\nomenclature{2D}{Two-dimensional}
\nomenclature{$\alpha$}{Thermal diffusivity $\left(\mathrm{m}^2 \mathrm{~s}^{-1}\right)$}
\nomenclature{$\rho$}{Density $\left(\mathrm{kg/m^3} \right)$}
\nomenclature{$g$}{Acceleration due to gravity $\left(\mathrm{m\ } \mathrm{s}^{-2}\right)$}
\nomenclature{$\theta$}{Temperature (dimensionless)}
\nomenclature{$p$}{Pressure (dimensionless)}
\nomenclature{$\delta_e$}{Relative error}
\nomenclature{$C_p$}{Specific heat capacity}
\nomenclature{$Ra$}{Rayleigh number}
\nomenclature{$Pr$}{Prandtl number}
\nomenclature{$\tau$}{Dimensionless time}
\nomenclature{$L$}{Cavity length ($m$)}
\nomenclature{$\phi$}{Volumev fraction of nanoparticles  ($m$)}
\nomenclature{$T$}{Temperature $(\mathrm{K})$}
\nomenclature{$x, y, z$}{Cartesian coordinates}
\nomenclature{$u, v, w$}{Velocity components (dimensionless) in $x, y, z$ directions}
\nomenclature{$Nu$}{Nusselt number}

\nomenclature{$\beta$}{Thermal expansion coefficient $\left(\mathrm{K}^{-1}\right)$}
\nomenclature{$\mu$ }{Dynamic viscosity $\left(\mathrm{kg\ } \mathrm{m}^{-1} \mathrm{~s}^{-1}\right)$}
\nomenclature[Sp]{$L$}{Local}
\nomenclature[Sp]{$Avg$}{Average}
\nomenclature[sp]{$nf$}{Nanofluid}
\nomenclature[sp]{$bf$}{Base fluid}
\nomenclature[sp]{$np$}{Nanoparticles}
\nomenclature[sp]{$s$}{Solid fin}
\nomenclature[sp]{$T$}{Total}

\printnomenclature

\section{Introduction}
In recent years, enhancing heat transfer has become a key focus for engineers and scientists due to its wide range of applications, such as improving energy efficiency in industrial processes, enhancing thermal management in electronics, and optimizing renewable energy systems. In this context, natural convection has been recognized as the most efficient mode of heat transfer. Various methods for enhancing heat transfer, including placing fins, using porous media, and other techniques, have been explored. The relatively poor thermal conductivity of frequently used fluids such as oil, air, and water presents a major challenge in improving heat transfer in engineering systems. To address this limitation, there is a pressing need to develop advanced heat transfer fluids with much better thermal conductivity. One approach is through the use of nanofluids \cite{Choi_1995,Lee_1999,Eastman_2001}. A nanofluid is a dilute suspension of solid nanoparticles, typically less than 100 nm in size, dispersed in a base fluid like water, oil, or ethylene glycol. Nanofluids have better thermal properties compared to base fluids or conventional conventional particle-fluid mixtures. They are stable, cause minimal pressure drops, and can travel through nano-channels. Consequently, nanofluids are used in various scientific and biological applications, including blood clotting, drug and gene delivery, cell transport in arteries and veins, and the manufacturing of nanocomposites. The addition of nanoparticles to these fluids significantly enhances their thermal conductivity. In recent years, there have been numerous studies on convective heat transfer in nanofluids.\\
Initially, Khanafer et al. \cite{Khanafer_2003} conducted a two-dimensional (2D) numerical analysis on the natural convection of a nanofluid contained within a differentially heated enclosure by using the finite-volume approach. They utilized $Al_2O_3$ based nanofluid and observed a significant enhancement in heat transfer rates at any given Grashof number. Moreover, the results demonstrate that increasing the volume percentage of nanoparticles improves heat transmission rates. Similar findings are reported in the study by Jou and Tzeng \cite{Jou_2006}, who conducted a numerical investigation of free convection in differentially heated two-dimensional rectangular cavities filled with a nanofluid. 
Oztop and Abu-nada \cite{Oztop_2008} performed computational simulations to analyze the natural convection behavior of nanofluids within partially heated rectangular cavities. The cavities featured a localized heater on one vertical wall, with the opposite vertical wall kept cold and the horizontal walls insulated. Their findings show that greater volume percentages of nanoparticles and taller heaters lead to an increase in the average Nusselt number. Aminossadati and Ghasemi \cite{Aminossadati_2009} conducted a 2D numerical investigation on the free/natural convection of nanofluid within a square cavity. Two vertical walls and the top horizontal wall cooled the cavity, while the bottom horizontal wall had a constant heat flux heater. They explored the influences of the volume fraction of nanoparticles, Rayleigh number, type of nanofluids, and the size and placement of the heater. 
Abu-nada and Oztop \cite{Abu_2009} investigated how the inclination angle of a 2D square cavity affects the natural convection of Cu-water nanofluid within it. They noted that the inclination angle could serve as a control parameter for heat transfer and fluid flow within the cavity. Qin et al. \cite{Qin_2014} developed a more efficient high-order compact difference method to investigate natural double diffusion in a 2D rectangular vertical cavity. The cavity contained a binary mixture with horizontal temperature and concentration gradients. They identified a bifurcation pattern that varied with the Prandtl number, and observed two distinct types of periodic flow patterns depending on the Lewis number.
Kumar et al. \cite{Kumar_2010} conducted an analysis of the natural convection of nanofluid in a 2D square cavity with differential heating, employing a single-phase thermal dispersion model. Using this model, simulations showed increased local thermal conductivity at maximum velocity sites. Additionally, their findings demonstrated that the average Nusselt number rose with the solid volume fraction.
Mahmoodi \cite{Mahmoodi_2011} investigated the 2D natural convection of Cu-water nanofluid inside an L-shaped cavity. He also got a similar relation between the solid volume fraction of the nanofluids and the average Nusselt number.
Numerous other studies have shown conflicting findings, indicating a decrease in heat transfer efficiency when using nanofluids. 
For example, Santra et al. \cite{Santra_2008} carried out a numerical investigation to analyze the influence of copper-water nanofluid on natural convection heat transfer in a differentially heated 2D square cavity, considering non-Newtonian behavior of nanofluids. They found that increasing the concentration of nanoparticles led to a decrease in heat transfer at a specific Rayleigh number. Recently, researchers have concentrated on improving the heat transfer of nanofluids natural convection in enclosures by integrating fins and baffles into the wall structures. 
Zemani et al. \cite{Zemani_2023} investigated the impact of baffle shape and
aspect ratio on 2D natural convection processes within a U-shaped cavity filled with CuO water-based nanofluid, focusing on a T-shaped insulating baffle attached to the cold wall. They observed that the T-shaped baffle significantly improves heat transfer rate. Rahmati et al. \cite{Rahmati_2018}
explore the natural convection of water-$TiO_2$ nanofluid in a 2D square cavity with a heated (constant heat) obstacle by using the lattice Boltzmann method. Their study demonstrated that increasing the barrier size to 0.5$\times$L greatly increased the average Nusselt number, but increasing the size to 0.7$\times$L resulted in a considerable decrease. Jamesahar et al. \cite{Jamesahar_2020} investigated the impact of mixed convection of nanofluids in a 2D cavity with flexible fins. They employed the finite element method and arbitrary Lagrangian–Eulerian technique to solve the governing equations. Bendara et al. \cite{Bendaraa_2019} investigated heat transfer characteristics of CuO-water nanofluid inside a 2D square enclosure with fins attached to the cold, hot, and adiabatic walls with the finite difference method using the upwind scheme. Ultimately, they found that attaching fins to adiabatic and cold walls enhances heat transfer performance, whereas placing fins on hot walls reduces it. 

However, it's noteworthy that most of these studies have focused on 2D configurations due to the high complexity involved in 3D scenarios. Recognizing that real fluid flows are inherently three-dimensional, there have been some efforts to extend simulations into 3D natural convection of nanofluids as well. Ravnik et al. \cite{Ravnik_2010} investigated the flow and heat transfer characteristics of nanofluids in 3D enclosed cavities under natural convection. They used a three-dimensional boundary element method to solve the velocity-vorticity formulation of the Navier-Stokes equations. They found that water-based nanofluids enhance heat transfer, with the largest improvement observed when conduction dominates, due to the high thermal conductivity of nanofluid. Rahimi et al. \cite{Rahimi_2017} conducted both experimental and numerical investigations on natural convection heat transfer performance in a 3D cuboid enclosure filled with Double-Walled Carbon Nanotubes (DWCNTs)-water nanofluids. The experiments were conducted with four different temperature differences between the hot and cold side walls and six different solid volume fractions of nanofluids and a finite volume approach was used for the numerical solution. They found that varying the temperature difference significantly impacts the temperature distribution, whereas increasing the solid volume fraction has minimal effect, especially in the enclosure's core region. Using the SIMPLER method, Ramón et al. \cite{Ramón_2007} explored the 3D natural convection of air within a cubical enclosure featuring a fin on the heated wall. By varying the solid-to-fluid thermal conductivity ratio ($R_k$) and the fin width, they found that higher $R_k$ values led to a substantial increase in the contribution from the fin faces, resulting in heat transfer improvements exceeding 20\%. These improvements are greater than those typically observed in most two-dimensional studies. Salari et al. \cite{Salari_2017} investigated the entropy generation and natural convection in a 3D cuboid enclosure filled with Multi-Walled Carbon Nanotubes (MWCNTs)-water nanofluid and air, which included a free surface. The vorticity-vector potential formulation of the Navier-Stokes equations is employed to remove the pressure term in 3D geometries. The equations are then solved using the control volume finite difference technique. Their study analyzed the effects of various key parameters, including solid volume fraction of the nanofluid, aspect ratio, and Rayleigh number on fluid flow and heat transfer efficiency. Rahimi et al. \cite{Rahimi_2019} explored the 3D natural convection and entropy generation in a cuboid cavity containing CuO-water nanofluids, employing the lattice Boltzmann method in their study. They found that the Nusselt number increases as both the Rayleigh number and the solid volume fraction of nanofluids increase. Sannad et al. \cite{Sannad_2020} studied the natural convection of nanofluids within a 3D cavity using the finite volume element method. Their research explored the fluid flow and heat transfer characteristics across various governing parameters, including $Ra$, volume fraction ($\phi$), and nanofluid type. The findings revealed a significant enhancement in heat transfer with increasing $Ra$ and $\phi$. Among the different nanofluids analyzed, copper-based nanofluid exhibited superior thermal performance, highlighting its potential for optimizing heat transfer efficiency. Kuharat et al. \cite{Kuharat_2020} performed finite volume based fluid flow simulations to analyze steady-state natural convection in a 3D tilted prismatic solar enclosure using gold-water nanofluids. Notable changes in vortex structure and temperature distribution are observed with variations in Rayleigh number, tilt angle, volume fraction, and aspect ratio. Hashemi et al. \cite{Hashemi_2024} studied the heat transfer enhancement under natural convection in a 3D cavity by placing an oscillating plate in various locations on the top of the cavity to examine its effect on heat transfer performance. This study utilizes two-way FSI analysis and the finite volume method (FVM).

A review of the literature reveals that most nanofluid convection studies are confined to 2D, highlighting a significant gap in 3D analyses. Additionally, there is a noticeable lack of higher-order accurate finite difference methods for nanofluid convection. Realistic and dependable simulations require both 3D modeling and higher-order accuracy. So, there is a continuous demand for higher-order accurate methods to study 3D nanofluid convection, creating a valuable opportunity for improved accuracy and advancements in this field. Although extensive research has focused on solid fins with insulated or constant heat sources for further heat transfer enhancement, there is a notable lack of studies on heat-conducting fins with nanofluid. Conducting fins offers the potential to greatly enhance thermal efficiency by facilitating active heat transfer and increasing the effective heat exchange surface. Investigating this approach is crucial for optimizing heat management in critical applications such as electronics cooling, where efficient thermal dissipation is vital for performance and longevity.

To address these gaps, we extend the higher-order super-compact finite difference scheme to investigate the impact of conducting fins on the natural convection of nanofluids. First, Kalita \cite{Kalita_2014} introduced the higher-order super-compact scheme and used it for a simple lid-driven cavity problem without considering heat or energy transfer. Recently, Punia and Ray \cite{Punia_2024} extended the HOSC scheme to study natural convection and entropy generation in a 3D air-filled cavity. They found that the method can accurately capture the heat transfer phenomena with high precision. Due to the high importance of enhancing heat transfer through nanofluid convection, we are now extending this scheme to study the 3D natural convection of nanofluids. This scheme has fourth-order accuracy in spatial variables and second-order accuracy in time variables. Also, It uses just seven directly adjacent grid points at the $(n+1)^{th}$ time level to do finite difference discretization. Apart from the new scheme, we explored two cases: first, the natural convection of a nanofluid in a cubic cavity, and second, the impact of adding two conducting fins on the heated wall to analyze the further enhancement of heat transfer within the nanofluid filled 3D cavity. Our findings revealed several novel phenomena, which are discussed in the results and discussion section, providing intriguing and valuable insights.

\label{sec:Introduction}
\section{Problem Definition and Solution Methodology}
\label{sec:Problem Description and Discretization of Governing Equations}
\subsection{Problem Definition}
\begin{figure}
    \centering
    \includegraphics[width=\textwidth]{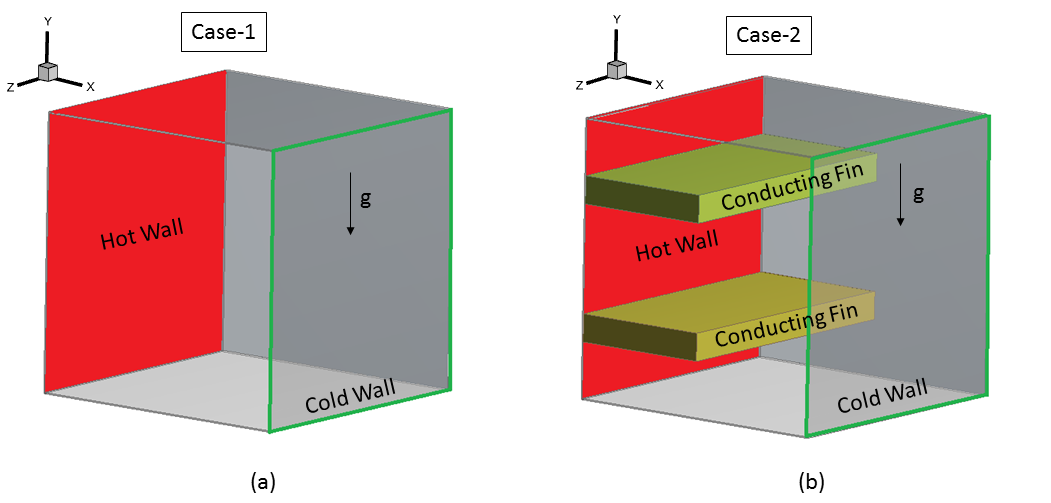}
    \caption{Schematic diagram of the configuration in the 3D cavity: (a) Case 1 (b) Case 2}
    \label{fig:Sche_diag}
\end{figure}

This work numerically investigates the natural convection flow within a closed 3D cubic cavity filled with a water-based nanofluid with $Cu$ (copper) nanoparticles.
The right wall at $x=1$ is kept cold, the left wall at $x=0$ is heated, and the remaining walls are adiabatic. We investigate two cases: first, the natural convection of nanofluid within a 3D cubic cavity, and second, inclusion of two aluminum conducting fins, each with a length of $L/2$ and a width of $L/10$, attached to the heated wall of the cavity. Here, $L$ represents the length of the cavity. The schematic diagrams for both cases are presented in Figure \ref{fig:Sche_diag}.
The nanofluid in the enclosure is considered incompressible and Newtonian, with constant material properties for the simulation. However, the viscosity and thermal conductivity of the nanofluid vary with the nanoparticle concentration. The governing equations are discretized using the higher-order super-compact (HOSC) scheme, with the Boussinesq approximation applied to account for density variations due to temperature changes. The physical properties of nanofluid are defined using the following equations:

\begin{itemize}
\item The effective heat capacitance of nanofluid \cite{Sannad_2019}:
$$
\left(\rho C_p\right)_{n f}= (1-\phi)\left(\rho C_p\right)_{bf}+\phi\left(\rho C_p\right)_{n p}.
$$
\item Density \cite{Sannad_2019}:
$$
\rho_{n f}=\phi \rho_{n p} - (\phi - 1) \rho_{bf}.
$$

 \item Thermal expansion coefficient \cite{Sannad_2019}:
$$
\beta_{n f}=\phi \beta_{n p}+(1-\phi) \beta_{bf}.
$$
\item The effective viscosity of the nanofluid is described by the Brinkman model \cite{Brinkman_1952}:
$$
\mu_{n f}=\mu_{bf}(1-\phi)^{-2.5} .
$$
\item Thermal diffusivity of nanofluids \cite{Sannad_2020}:
$$
\alpha_{n f}= (\left(\rho C_p\right)_{n f})^{-1}  k_{n f} .
$$
\item The Maxwell model expresses the thermal conductivity of the nanofluids \cite{Maxwell_1881}:
$$
\frac{k_{n f}}{k_{bf}}=\frac{2 k_{bf}-\left[2\phi\left(k_{bf}-k_{n p}\right) \right]+k_{n p}}{2 k_{bf}+\left[\phi\left(k_{bf}-k_{n p}\right)\right]+k_{n p}},
$$
Here, $k_{n p}, k_{bf}$, and $k_{nf}$ represent the thermal conductivities of solid nanoparticles, base fluid, and nanofluid, respectively. Table \ref{thermophysical_properties_Table} lists the thermophysical characteristics of both the base fluid and the nanoparticles. 
\end{itemize}

The governing equations for continuity, momentum, and energy conservation, expressed in Cartesian coordinates, are presented in the following dimensionless form \cite{Sannad_2020}:

\begin{equation} \label{Main_governing_eq_5} 
\frac{\partial w}{\partial z}+\frac{\partial v}{\partial y}+\frac{\partial u}{\partial x}=0 
\end{equation}
\begin{equation}\label{Main_governing_eq_1}
\frac{\partial u}{\partial \tau}+w \frac{\partial u}{\partial z}+v \frac{\partial u}{\partial y}+u \frac{\partial u}{\partial x}+\frac{\partial p}{\partial x} =\frac{\mu_{nf}}{\alpha_{bf}\rho_{nf}} \left[\frac{\partial^2 u}{\partial z^2}+\frac{\partial^2 u}{\partial y^2}+\frac{\partial^2 u}{\partial x^2}\right] 
\end{equation}
\begin{equation}\label{Main_governing_eq_2}
\frac{\partial v}{\partial \tau}+w \frac{\partial v}{\partial z}+v \frac{\partial v}{\partial y}+u \frac{\partial v}{\partial x}+\frac{\partial p}{\partial y}=\frac{\mu_{nf}}{\alpha_{bf}\rho_{nf}} \left[\frac{\partial^2 v}{\partial z^2}+\frac{\partial^2 v}{\partial y^2}+\frac{\partial^2 v}{\partial x^2}\right] +\frac{\rho_{bf}\beta_{nf}}{\beta_{bf}\rho_{nf}}* R a *\ P r *\ \theta 
\end{equation}

\begin{equation}\label{Main_governing_eq_3}
\frac{\partial w}{\partial \tau}+w \frac{\partial w}{\partial z}+v \frac{\partial w}{\partial y}+u \frac{\partial w}{\partial x}+\frac{\partial p}{\partial z}=\frac{\mu_{nf}}{\alpha_{bf}\rho_{nf}} \left[\frac{\partial^2 w}{\partial z^2}+\frac{\partial^2 w}{\partial y^2}+\frac{\partial^2 w}{\partial x^2}\right]
\end{equation}

\begin{equation}\label{Main_governing_eq_4}
\frac{\partial \theta}{\partial \tau}+ w \frac{\partial \theta}{\partial z}+v \frac{\partial \theta}{\partial y}+u \frac{\partial \theta}{\partial x}=\frac{\alpha_{nf}}{\alpha_{bf}} \left[\frac{\partial^2 \theta}{\partial z^2}+\frac{\partial^2 \theta}{\partial y^2}+\frac{\partial^2 \theta}{\partial x^2}\right]
\end{equation}
Inside the conducting fin, the energy equation (heat conduction) is
\begin{equation}\label{Main_governing_eq_4_fin}
\frac{\partial \theta}{\partial \tau}=\frac{\partial^2 \theta}{\partial z^2}+\frac{\partial^2 \theta}{\partial y^2}+\frac{\partial^2 \theta}{\partial x^2}
\end{equation}
The dimensionless expressions for the initial and boundary conditions are as follows:
$$
\begin{array}{lll}
\tau=0: \quad 0 \leqslant x, y, z \leqslant 1; \quad \theta=0;  \quad  w, v, u=0   ; \\
\tau>0:\quad x=0; \quad \theta=1; \quad  w, v, u=0 ;   \\
\quad \quad \quad \quad z=0,1; \quad \frac{\partial \theta}{\partial z}=0 ; \quad w, v, u=0 ;\\
\quad \quad \quad \quad y=0,1 \quad \frac{\partial \theta}{\partial y}=0 ; \quad  w, v, u=0 ;\\
\quad { }\quad \quad \quad x=1; \quad \theta=0; \quad  w, v, u=0 ;   \\
\end{array}
$$
Case 1 and Case 2 have the same initial and boundary conditions, with the exception of the additional boundary conditions for the conducting fins in Case 2. The base of the conducting fin (\(x = 0\)) is at \(\theta = 1\),  and the other faces of the fins are subject to the following boundary constraints \cite{Ramón_2007}:\\
$$\frac{\partial \theta_s}{\partial n} = \frac{k_{nf}}{k_s}\frac{\partial \theta_{nf}}{\partial n} $$
where $n$ represents normal to the surface boundary and subscripts $nf$ and $s$ indicate that the gradients are evaluated on the nanofluid and solid sides of the interface, respectively. We use aluminium conducting fins because of their excellent heat conductivity and cost-effectiveness. Aluminum has a thermal conductivity ($k$) of 237 $W.m^{-1}.K^{-1}$.
{\small\begin{table}[htbp]
\caption{\small Thermophysical properties of pure fluid (water) and copper nanoparticles}\label{thermophysical_properties_Table}
\centering
 \begin{tabular}{cccccc}   \hline 
 & $C_p$    &    $\beta$   &  $\rho$  &  $k$   \\ \hline 
Pure water & 4179   &   21 $\times$ $10^{-5}$   & 997.1   & 0.613   \\
$Cu$ & 385 &    1.67 $\times$ $10^{-5}$   &  8933  &   400 \\
\hline 
\hline
 \end{tabular}
\end{table}
}
\subsection{Discretization of Governing Equations with HOSC Scheme}
\begin{figure}
    \centering
    \includegraphics[width=\textwidth]{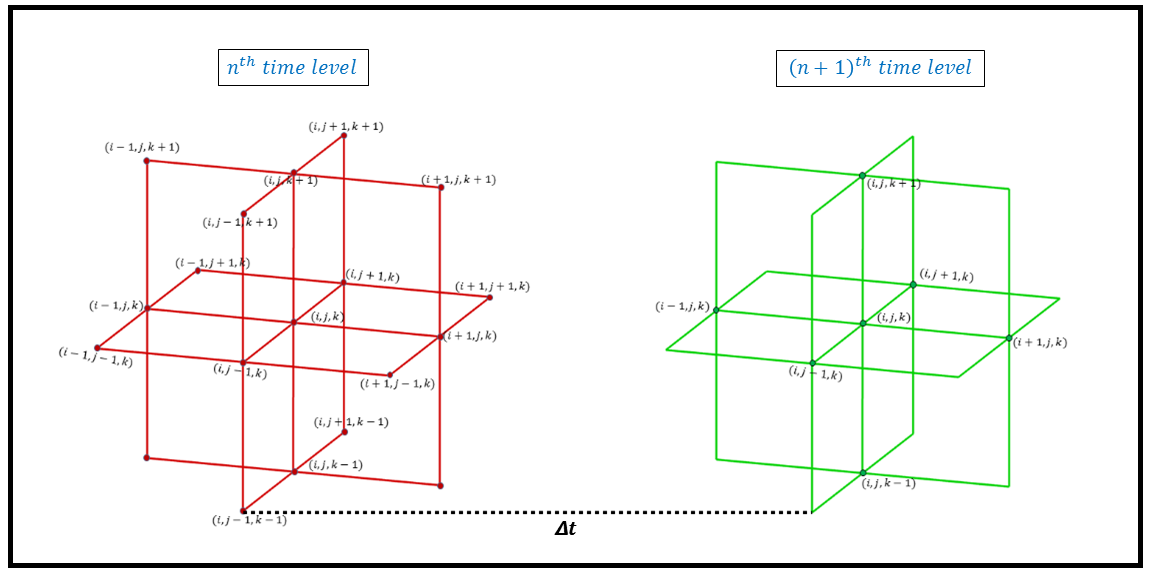}
    \caption{The super-compact unsteady stencil}
    \label{fig:stencil}
\end{figure}

The continuity, momentum, and energy equations govern the fluid flow and heat transfer of nanofluids within the three-dimensional cavity.
These equations are first non-dimensionalized as previously mentioned (equations (\ref{Main_governing_eq_5})-(\ref{Main_governing_eq_4_fin})) and then discretized on a structured grid. The inclusion of diffusion terms in the continuity associated with nanofluids significantly complicates the governing equations compared to conventional natural convection with simple fluids. These additional terms introduce complexities in the analysis, as they account for the enhanced thermal conductivity and modified flow characteristics inherent to nanofluid. Consequently, the system requires more sophisticated numerical approaches to accurately capture the intricate interactions between heat transfer and fluid dynamics. In this section, we shall discretise the governing equations using the new higher-order super-compact (HOSC) method. We consider the transient 3D convection-diffusion-reaction equation containing a transport variable, designated as ''$\xi$" inside a continuous domain. It is governed by variable coefficients and is stated by the equation below:
\begin{equation}
\begin{aligned}
& L\frac{\partial \xi}{\partial\tau}+K^{*}(x, y, z,\tau) \frac{\partial \xi}{\partial x}+M^{*}(x, y, z,\tau) \frac{\partial \xi}{\partial y}+N^{*}(x, y, z,\tau) \frac{\partial \xi}{\partial z}+O^{*}(x, y, z,\tau) \xi \\
& \quad=\nabla^2 \xi+P^{*}(x, y, z,\tau),
\end{aligned}\label{eq1}
\end{equation}
Here, $L$ is a constant, while $K^{*}, M^{*}, N^{*}, O^{*}$ and $P^{*}$ are functions of $\tau, x, y$, and  $z$. This equation describes the convection-diffusion process for different fluid variables, such as vorticity, streamfunction, energy, mass, and, heat within a continuous domain. This equation can accurately describe both the momentum and energy equations by selecting appropriate values for $L, O^{*}, N^{*}, M^{*}, K^{*}$, and $P^{*}$. Hence, this equation provides a complete foundation for expressing a large range of fluid dynamics phenomena in a single mathematical framework. To define the problem clearly and ensure its physical relevance, suitable boundary conditions need to be set for the domain. We apply a uniform mesh to discretize the cubic domain. With this grid setup, we first use the Forward-Time Centered-Space (FTCS) approach to approximate equation (\ref{eq1}) at the generic mesh point \((i, j, k)\). The FTCS approximation for equation (\ref{eq1}) at the node \((i, j, k)\) is as follows:
\begin{equation}
\begin{aligned}
\left(L \delta_\tau^{+}+O^{*}+N^{*}\delta_z+M^{*}\delta_y+K^{*} \delta_x-\delta_z^2-\delta_y^2-\delta_x^2\right) \xi_{i j k}-\varepsilon_{i j k}=P^{*}_{i j k},
\end{aligned}\label{eq2}
\end{equation}
The above equation includes key operators: \(\delta_z\), \(\delta_z^2\), \(\delta_x\), \(\delta_x^2\), \(\delta_y\), \(\delta_y^2\), and \(\delta_\tau^+\). These operators represent first- and second-order central differences in the \(z\)-, \(x\)-, and \(y\)-directions, and a first-order forward difference in the temporal direction, respectively. \(\xi_{ijk}\) represents the function value of \(\xi\) at the 3D grid point \((x_i, y_j, z_k)\). In numerical analysis, truncation error occurs when a finite number of terms are used to approximate an infinite process (Taylor series). It is the distinction between an exact mathematical quantity and a numerical approximation. Therefore, here, the truncation error \(\varepsilon_{ijk}\) for a uniform time step \(\Delta \tau\) can be expressed as:
\begin{equation}
\begin{aligned}
\varepsilon_{i j k}= & {\left[-\frac{h^2}{12}\left(\frac{\partial^4 \xi}{\partial x^4}-2 K^{*} \frac{\partial^3 \xi}{\partial x^3}\right)-\frac{k^2}{12}\left(\frac{\partial^4 \xi}{\partial y^4} - 2 M^{*} \frac{\partial^3 \xi}{\partial y^3}\right)\right.} \\
& \left.-\frac{l^2}{12}\left(\frac{\partial^4 \xi}{\partial z^4}-2 N^{*} \frac{\partial^3 \xi}{\partial z^3}\right)+L \frac{\Delta\tau}{2} \frac{\partial^2 \xi}{\partial\tau^2}\right]_{i j k}+O\left(\Delta\tau^2, h^4, k^4, l^4\right).
\end{aligned}\label{eq3}
\end{equation}
In order to attain a higher accuracy in space (fourth-order accuracy) and time (second-order accuracy) for equation (\ref{eq1}) in a compact way, the leading term's derivatives in equation (\ref{eq3}) are compactly approximated \cite{MacKinnon_1991, Spotz_1995}, resulting in an approach with minimized truncation error. Compactness in numerical schemes involves using a smaller, more efficient set of grid points to achieve high-order accuracy. For this purpose, the main equation (\ref{eq1}) is considered an extra relation, allowing for the derivation of higher-order derivatives ($3^{rd}$ and $4^{th}$). In other words, higher-order derivatives of equation (\ref{eq3}) are derived directly from the main equation (\ref{eq1}). For the transport variable \(\xi\), the forward temporal difference approach is applied, whereas the backward difference technique is applied to the variables \(L\), \(O^{*}\), \(M^{*}\), \(K^{*}\), \(N^{*}\), and \(P^{*}\) \cite{Kalita_2014}.
This enables us the final time derivative term of the equation (\ref{eq3}) to be expressed as:
\begin{equation}
\begin{aligned}
\left.L \frac{\partial^2 \xi}{\partial t^2}\right|_{i j k}= & \left(\delta_z^2+\delta_y^2+\delta_x^2-O^{*}_{i j k}-K^{*}_{i j k} \delta_x-M^{*}_{i j k} \delta_y-N^{*}_{i j k} \delta_z\right) \delta_\tau^{+} \xi_{i j k} \\
& -\left(\delta_\tau^{-} K^{*}_{i j k} \delta_x+\delta_\tau^{-} M^{*}_{i j k} \delta_y+\delta_\tau^{-} K^{*}_{i j k}+\delta_\tau^{-} N^{*}_{i j k} \delta_z +\delta_\tau^{-} P^{*}_{i j k}\right) \xi_{i j k}\\
&+O\left(\Delta\tau, h^2, k^2, l^2\right),
\end{aligned} \label{eq4}
\end{equation}
The operator \(\delta_\tau^{-}\) represents a first-order backward difference, while \(\delta_\tau^{+}\) denotes a first-order forward difference with respect to the time. Similarly, equivalent approximations can also be applied to the spatial derivatives. By replacing these derivatives from equation (\ref{eq4}) and similar expressions into equation (\ref{eq3}), and updating \(\varepsilon_{ijk}\) in equation (\ref{eq2}), we obtain an approximation of order \(O\left(\Delta \tau^2, h^4, k^4, l^4\right)\) for the main governing equation (\ref{eq1}). This can be expressed as:
$$
\begin{aligned}
L[1 & +\left(\frac{h^2}{12}-\frac{\Delta\tau}{2 L}\right)\left(\delta_x^2-K^{*}_{i j k} \delta_x\right)+\left(\frac{k^2}{12}-\frac{\Delta\tau}{2 L}\right)\left(\delta_y^2-M^{*}_{i j k} \delta_y\right) \\
& \left.+\left(\frac{l^2}{12}-\frac{\Delta\tau}{2 L}\right)\left(\delta_z^2-N^{*}_{i j k} \delta_z\right)+\frac{\Delta\tau}{2 L} O^{*}_{i j k}\right] \delta_\tau^{+} \xi_{i j k} \\
& +\left(-\alpha_{i j k} \delta_x^2-\beta_{i j k} \delta_y^2-\gamma_{i j k} \delta_z^2+A1_{i j k} \delta_x+B1_{i j k} \delta_y+C1_{i j k} \delta_z+D1_{i j k}\right) \xi_{i j k} \\
& -\frac{h^2+k^2}{12}\left(\delta_x^2 \delta_y^2-K^{*}_{i j} \delta_x \delta_y^2-M^{*}_{i j k} \delta_x^2 \delta_y-p1_{i j k} \delta_x \delta_y\right) \xi_{i j k}
\end{aligned}
$$
\begin{equation}
\begin{aligned}
& -\frac{k^2+l^2}{12}\left(\delta_y^2 \delta_z^2-M^{*}_{i j k} \delta_y \delta_z^2-N^{*}_{i j k} \delta_y^2 \delta_z-q1_{i j k} \delta_y \delta_z\right) \xi_{i j k} \\
& -\frac{l^2+h^2}{12}\left(\delta_z^2 \delta_x^2-N^{*}_{i j k} \delta_z \delta_x^2-K^{*}_{i j k} \delta_z^2 \delta_x-r1_{i j k} \delta_z \delta_x\right) \xi_{i j k} \\
= &  R_{i j k} .
\end{aligned}\label{eq5}
\end{equation}
The coefficients $A1_{i j k}, B1_{i j k}, C1_{i j k}, D1_{i j k}, R_{i j k}, \beta_{i j k}, \alpha_{i j k}, \gamma_{i j k}, p1_{i j k}, r1_{i j k}$ and $q1_{i j k}$ are as follows:
$$
\begin{aligned}
& A1_{i j k}=\left[\frac{h^2}{12}\left(\delta_x^2-{K^{*}}_{i j k} \delta_x\right)+\frac{k^2}{12}\left(\delta_y^2-{M^{*}}_{i j k} \delta_y\right)+\frac{l^2}{12}\left(\delta_z^2-{N^{*}}_{i j k} \delta_z\right)+\frac{\Delta\tau}{2} \delta_\tau^{-}+1\right] {K^{*}}_{i j k} \\
& -\frac{h^2}{12}\left({K^{*}}_{i j k}-2 \delta_x\right) {O^{*}}_{i j k}, \\
& B1_{i j k}=\left[\frac{h^2}{12}\left(\delta_x^2-{K^{*}}_{i j k} \delta_x\right)+\frac{k^2}{12}\left(\delta_y^2-{M^{*}}_{i j k} \delta_y\right)+\frac{l^2}{12}\left(\delta_z^2-{N^{*}}_{i j k} \delta_z\right)+\frac{\Delta\tau}{2} \delta_\tau^{-}+1\right] {M^{*}}_{i j k} \\
& -\frac{k^2}{12}\left({M^{*}}_{i j k}-2 \delta_y\right) {O^{*}}_{i j k} \text {, } \\
& C1_{i j k}=\left[\frac{h^2}{12}\left(\delta_x^2-{K^{*}}_{i j k} \delta_x\right)+\frac{k^2}{12}\left(\delta_y^2-{M^{*}}_{i j k} \delta_y\right)+\frac{l^2}{12}\left(\delta_z^2-{N^{*}}_{i j k} \delta_z\right)+\frac{\Delta\tau}{2} \delta_\tau^{-}+1\right] {N^{*}}_{i j k} \\
& -\frac{l^2}{12}\left({N^{*}}_{i j k}-2 \delta_z\right) {O^{*}}_{i j k} \text {, } \\
& D1_{i j k}=\left[\frac{h^2}{12}\left(\delta_x^2-{K^{*}}_{i j k} \delta_x\right)+\frac{k^2}{12}\left(\delta_y^2-{M^{*}}_{i j k} \delta_y\right)+\frac{l^2}{12}\left(\delta_z^2-{N^{*}}_{i j k} \delta_z\right)+\frac{\Delta\tau}{2} \delta_\tau^{-}+1\right] {O^{*}}_{i j k}, \\
& R_{i j k}=\left[\frac{h^2}{12}\left(\delta_x^2-{K^{*}}_{i j k} \delta_x\right)+\frac{k^2}{12}\left(\delta_y^2-{M^{*}}_{i j k} \delta_y\right)+\frac{l^2}{12}\left(\delta_z^2-{N^{*}}_{i j k} \delta_z\right)+\frac{\Delta\tau}{2} \delta_\tau^{-}+1\right] {P^{*}}_{i j k}, \\
& \alpha_{i j k}=\frac{h^2}{12}\left({K^{*}}_{i j k}^2-{O^{*}}_{i j k}-2 \delta_x {K^{*}}_{i j k}\right)+1 \text {, } \\
& \gamma_{i j k}=\frac{l^2}{12}\left({N^{*}}_{i j k}^2-{O^{*}}_{i j k}-2 \delta_z {N^{*}}_{i j k}\right)+1\text {, } \\
& \beta_{i j k}=\frac{k^2}{12}\left({M^{*}}_{i j k}^2-{O^{*}}_{i j k}-2 \delta_y {M^{*}}_{i j k}\right)+1 \text {, } \\
& p1_{i j k}=-{K^{*}}_{i j k} {M^{*}}_{i j k}+\frac{2}{h^2+k^2}\left(k^2 \delta_y {K^{*}}_{i j k}+h^2 \delta_x {M^{*}}_{i j k}\right) \text {, } \\
& q1_{i j k}=-{M^{*}}_{i j k} {N^{*}}_{i j k}+\frac{2}{k^2+l^2}\left(l^2 \delta_z {M^{*}}_{i j k}+k^2 \delta_y {N^{*}}_{i j k}\right) \text {, } \\
& r1_{i j k}=-{N^{*}}_{i j k} {K^{*}}_{i j k}+\frac{2}{l^2+h^2}\left(h^2 \delta_x {N^{*}}_{i j k}+l^2 \delta_z {K^{*}}_{i j k}\right). 
&
\end{aligned}
$$
From equation (\ref{eq5}), we derive an implicit finite difference scheme having second and fourth-order accuracy \(O(\tau^2, h^4, k^4, l^4)\) in time and space, respectively. This high accuracy is attained using \((19,7)\) stencil, as illustrated in Figure \ref{fig:stencil}.

This method achieves a compact seven-point stencil over the \((n + 1)^{\text{th}}\) time level, significantly minimizing computing costs. Remarkably, even for the 3D case, this method only needs 7 points stencil at the \((n + 1)^{\text{th}}\) time level, whereas many high-order compact methods for 2D convection-diffusion equations \cite{Kalita_2002,Ray_2017} normally demand 9 points stencil at the same time level.
This approach significantly decreases the overall number of points necessary for approximation by minimizing the demand for numerous corner points, thereby improving computational efficiency. Reference \cite{Kalita_2014, Punia_2024} provides further more information on the higher-order super-compact scheme and coefficients.
The coefficient matrix for equation (\ref{eq5}) forms an asymmetric sparse matrix without diagonal dominance. As a result, standard iterative techniques such as Gauss-Seidel and SOR are not successful in this scenario. Hence, a Hybrid Biconjugate-Gradient Stabilized method is used to solve the system of equations generated by Equation (\ref{eq5}) without the use of preconditioning \cite{Kalita_2014,Spotz_1995,Kelley_1995,Saad_2003}.\\
The initial step in solving the governing equations is to discretize the momentum equations (Eqs. (\ref{Main_governing_eq_1}),(\ref{Main_governing_eq_2}),(\ref{Main_governing_eq_3})) using the designated HOSC scheme. To solve these momentum equations, we apply the HOSC discretization (Eq. \ref{eq5}) by utilizing the coefficient values of equation (\ref{eq1}) provided in Table \ref{Coefficient_Table}.
Following the solution of the momentum equations, the pressure field ($p$) is calculated. Unlike velocity components, pressure lacks a direct evolution equation, making its handling one of the most complex aspects of 3D Navier-Stokes equations. In contrast, many 2D studies avoid the complexities of pressure management by employing the streamfunction-vorticity formulation. In our 3D study, we chose the modified compressibility approach given by Cortes and Miller \cite{Cortes_1994} due to its simplicity and efficiency, which offer a faster and easier implementation. In this approach, the continuity equation is modified and written as:
$$
p+\lambda \nabla \cdot \mathbf{u}=0 .
$$
The dilation parameter $D=u_x+v_y+w_z$ is computed at each time step once the pressure gradients have been calculated and the momentum equations solved. If the maximum value of $|D|$ is below a specified tolerance, the pressure is deemed sufficiently accurate. The process then proceeds to the subsequent time step and continues iteratively. However, if the highest value of $|D|$ exceeds the given tolerance level, pressure is corrected by using a pressure correction step:
$$
p^{\text {1}}=p^{\text {0}}-\lambda \nabla \cdot \mathbf{u} .
$$
where \( \lambda \) is the relaxation factor, \( p^{\text{1}} \) denotes the updated pressure, and \( p^{\text{0}} \) is the pressure from the previous pressure iteration. The operation is performed repeatedly until the highest value of \( |\nabla \cdot \mathbf{u}|_{\max} \), meets the tolerance limit.
Thus, this approach completes a one time iteration while ensuring that the continuity equation is also satisfied at each time level. The procedure is continued until a steady state is achieved.
{\small\begin{table}[htbp] \footnotesize
\caption{\small Equation Coefficients}\label{Coefficient_Table}
\centering
 \begin{tabular}{cccccccc}  \hline \hline
Equation & $\xi$ & $L$   &    $K^{*}$   &  $M^{*}$  &  $N^{*}$  & $P^{*}$   \\
$x$-momentum &$u$ & $\frac{\alpha_{bf}\rho_{nf}}{\mu_{nf}}$   &    $u\frac{\alpha_{bf}\rho_{nf}}{\mu_{nf}}$   &  $v\frac{\alpha_{bf}\rho_{nf}}{\mu_{nf}}$  &  $w\frac{\alpha_{bf}\rho_{nf}}{\mu_{nf}}$   & $-(\frac{\alpha_{bf}\rho_{nf}}{\mu_{nf}}) \cdot \frac{\partial p}{\partial x}$ \\

$y$-momentum  &$v$& $\frac{\alpha_{bf}\rho_{nf}}{\mu_{nf}}$  &    $u\frac{\alpha_{bf}\rho_{nf}}{\mu_{nf}}$   &  $v\frac{\alpha_{bf}\rho_{nf}}{\mu_{nf}}$  &  $w\frac{\alpha_{bf}\rho_{nf}}{\mu_{nf}}$ & $-(\frac{\alpha_{bf}\rho_{nf}}{\mu_{nf}}) \cdot (\frac{\partial p}{\partial y} + \frac{\rho_{bf}\beta_{nf}}{\beta_{bf}\rho_{nf}} \cdot Ra \cdot Pr \cdot \theta) $ \\

$z$-momentum  &$w$& $\frac{\alpha_{bf}\rho_{nf}}{\mu_{nf}}$   &    $u\frac{\alpha_{bf}\rho_{nf}}{\mu_{nf}}$   &  $v\frac{\alpha_{bf}\rho_{nf}}{\mu_{nf}}$  &  $w\frac{\alpha_{bf}\rho_{nf}}{\mu_{nf}}$  & $-(\frac{\alpha_{bf}\rho_{nf}}{\mu_{nf}}) \cdot \frac{\partial p}{\partial z}$ \\

Energy (Nanofluid) & $\theta$ & $\frac{\alpha_{bf}}{\alpha_{nf}} $  &    $u$   &  $v$  &  $w$ & $0$ \\
Energy (Fin)  & $\theta$ & 1  &  $0$    &  $0$ & 0 & $0$ \\

\hline \\
\hline
 \end{tabular}
\end{table}
}

\newpage
\section{Validation and Sensitivity Analysis of the Scheme}
\label{sec:SENSITIVITY TESTS AND SCHEME VALIDATION}
\subsection{Grid Independence Test}
To guarantee the accuracy and efficiency of our findings, we carried out a thorough grid independence study. This process involved adjusting the mesh resolution systematically while keeping all other simulation parameters unchanged, ensuring that our results are not affected by grid size. We assessed three different grid resolutions: $11\times11\times11$, $51\times51\times51$, and $91\times91\times91$, while keeping the volume fraction ($\phi$), $\Delta \tau$ and $Ra$ fixed at 0.04, 0.01, and $10^5$, respectively. Case 1 is chosen for the grid independence test. Table \ref{grid_independent_test} shows the velocity measurements at two key observation points, \((0.60, 0.60, 0.60)\) and \((0.75, 0.75, 0.75)\), located near the core area of the cavity, for each grid resolution at \(\tau = 25\) and \(\tau = 50\). This study demonstrates that the maximum relative difference between the velocity values at $91\times91\times91$ and $51\times51\times51$ grids is just $1.93\%$. This suggests that further increasing the grid resolution beyond \(51 \times 51 \times 51\) provides only minimal improvements in results. Consequently, the \(51 \times 51 \times 51\) mesh is sufficient for capturing the flow dynamics with precision. Therefore, we have chosen the \(51 \times 51 \times 51\) grid size for our computational analysis.

{\small\begin{table}[htbp]
\caption{\small Velocity measurements at two key observation points \((0.60, 0.60, 0.60)\) and \((0.75, 0.75, 0.75)\), located near the core area of the cavity, with \(Ra = 10^5\) and \(\Delta \tau = 0.01\), obtained using 3 distinct grid resolutions.
}\label{grid_independent_test}
\centering
 \begin{tabular}{cccccccc}  \hline \hline
Observation point & Grid resolution    &    $u$ & $\delta_e$(\%)  &  $v$ & $\delta_e$(\%)  &  $w$ & $\delta_e$(\%)    \\ \hline
\textbf{time = 25}\\\hline
(0.60, 0.60, 0.60) & (11 $\times$ 11 $\times$ 11)   &  -6.980 & -- &   4.665 & --   &   -0.771  & -- \\
  & (51 $\times$ 51 $\times$ 51)      &  -8.563  & 22.6 &    5.686 & 21.8 &   -0.876 & 13.6 \\
& (91 $\times$ 91 $\times$ 91)  &  -8.729 & 1.93 &  5.698 &  0.21 &   -0.878  & 0.22  \\ 
\hline
(0.75, 0.75, 0.75)  & (11 $\times$ 11 $\times$ 11)                        & 9.658 & -- &  8.022 & --   &   0.919  & -- \\
& (51 $\times$ 51 $\times$ 51)                       & 9.113 & 5.64 &  10.458  & 30.3 &   1.063 & 15.6\\
 & (91 $\times$ 91 $\times$ 91)  &  8.987 & 1.38 &  10.477  & 0.18 &   1.068  & 0.47 \\
\hline 
 \textbf{time = 50}   \\
\hline
(0.60, 0.60, 0.60)  & (11 $\times$ 11 $\times$ 11)                 &  -3.646 & --&   4.680 & --  &   -0.683  & -- \\
 & (51 $\times$ 51 $\times$ 51)                 &     -4.466 & 22.4 &  5.585  & 19.3 &  -0.573 & 16.1\\
& (91 $\times$ 91 $\times$ 91)     &  -4.489 & 0.51 &  5.543 & 0.75 &   -0.569  & 0.69 \\
\hline
(0.75, 0.75, 0.75)  & (11 $\times$ 11 $\times$ 11)                   & 12.994 & --&   8.259 & --   &   1.947  & -- \\
& (51 $\times$ 51 $\times$ 51)                      & 15.095 & 16.1  &  10.108 & 22.3 &   2.454 & 26.0\\
 & (91 $\times$ 91 $\times$ 91)  &  15.311 & 1.43 &  10.193  & 0.84 &   2.481   & 1.10  \\
 
\hline
\hline
 \end{tabular}
\end{table}
}

\subsection{Validation of the HOSC Scheme for Nanofluid Natural Convection}
%%%%%%%%%%%%%%%%%%%%%%%%%%%%%%%%%%%%%%%%%%%%%%%%%%%%%%%%%%%%%%%%%%%%%%%%%%%%%%
\begin{figure}
    \centering
    \includegraphics[width=\textwidth]{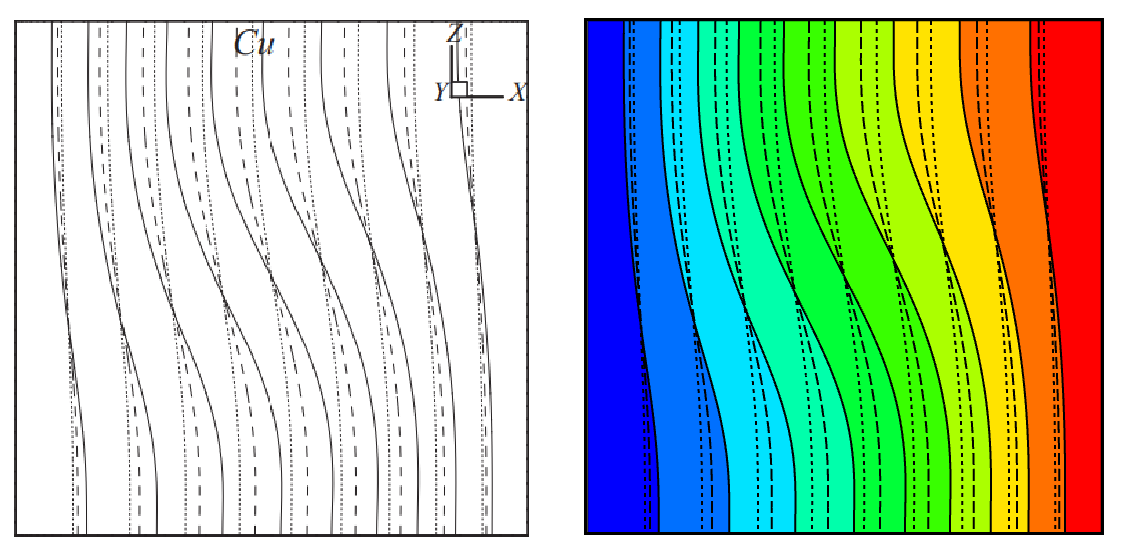}
    \caption{Comparison of isotherms on the symmetry plane ($y=0.5$) at $Ra=10^3$: Results of Ravnik et al. (3D) \cite{Ravnik_2010} vs. Present Study (Dotted line: $\phi=0.2$ nanofluid, Dashed line: $\phi=0.1$ nanofluid, Solid line: Pure Water)}
    \label{fig:3d_isotherm_comparison}
\end{figure}

%%%%%%%%%%%%%%%%%%%%%%%%%%%%%%%%%%%%%%%%%%%%%%%%%%%%%%%%%%%%%%%%%%%%%%%%%%%%%%%%%%%%%%%%%%%%%%%%%%%%%%%
\begin{figure}
    \centering
    \includegraphics[width=\textwidth]{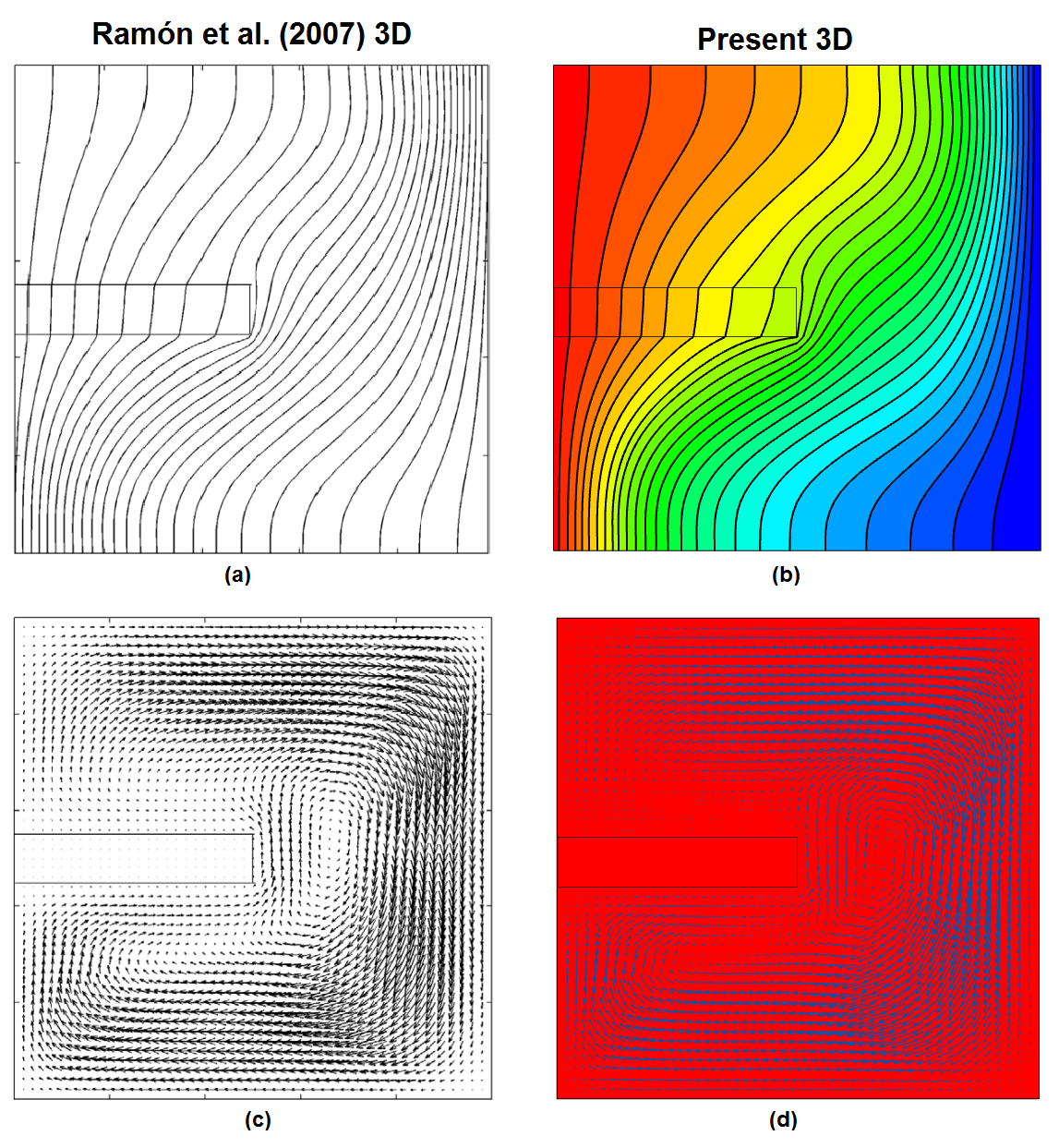}
    \caption{Comparison of temperature contours (a, b) and velocity vectors (c, d) on $z=0.5$, $R_k=10$, $Ra=10^4$: Results of Ramón et al. \cite{Ramón_2007} vs. Present work}
    \label{fig:fin_comparison}
\end{figure}

{\small\begin{table}[htbp]
\caption{\small Comparison of average Nusselt number at the hot wall for different volume fractions ($\phi = 0.0, 0.1, 0.2$) and $Ra$ $(10^3, 10^4, 10^5)$ values, incorporating the relative percentage error ($\delta_e$(\%)) with previously published results \cite{Ravnik_2010,Sannad_2020}.}\label{Average_Nusselt_Number_Comparison}
\centering
 \begin{tabular}{ccccccccccc}  \hline \hline
& &  $Ra$ & $\phi=0.0$  & $\delta_e$(\%)  &  $\phi=0.1$  & $\delta_e$(\%)  &  $\phi=0.2$  & $\delta_e$(\%)   \\ \hline 
& Present    &   $10^3$      &  1.075 & \textbf{--} &  1.365 & \textbf{--} &  1.759 & \textbf{--} &   \\
& Ravnik et al. \cite{Ravnik_2010} & $10^3$    &  1.071 & \textbf{0.37} & 1.363  & \textbf{0.14} &    1.758 & \textbf{0.05}\\
& Sannad et al. \cite{Sannad_2020}    &   $10^3$   &  1.080 & \textbf{0.46} &     1.371  & \textbf{0.43} &    1.767 & \textbf{0.45} \\ \hline 
& Present    &   $10^4$      &  2.105 & \textbf{--} &  2.264 & \textbf{--} &  2.404 & \textbf{--} &   \\
& Ravnik et al. \cite{Ravnik_2010} & $10^4$    &  2.078 & \textbf{1.29}  &  2.237  & \textbf{1.20} &   2.381 & \textbf{0.96}\\ 
& Sannad et al. \cite{Sannad_2020}    &   $10^3$   &  2.121 & \textbf{0.75} &     2.277  & \textbf{0.57} &    2.415 & \textbf{0.45} \\ \hline 
& Present    &   $10^5$      &  4.552 & \textbf{--} &  5.003 & \textbf{--} &  5.344 & \textbf{--} &   \\
& Ravnik et al. \cite{Ravnik_2010} & $10^5$    &   4.510 & \textbf{0.93}  &  4.946  & \textbf{1.15} &    5.278 & \textbf{1.25}\\
& Sannad et al. \cite{Sannad_2020}    &   $10^5$   &  4.672 & \textbf{2.56} &    5.095  & \textbf{1.80} &     5.409 & \textbf{1.20} \\
\hline\hline
 \end{tabular}
\end{table}
}
Validation of our code and scheme is critical to ensure that they accurately represent the intricate flow and heat transfer phenomena associated with 3D nanofluid natural convection. In this section, we systematically validate our in-house code and the HOSC scheme by comparing the results to existing benchmark results. Validation is carried out both quantitatively and qualitatively. The quantitative comparison is presented in Table \ref{Average_Nusselt_Number_Comparison}, where we compare our results (average Nusselt number) with the benchmark results from Ravnik et al. \cite{Ravnik_2010} and Sannad et al. \cite{Sannad_2020}. The average Nusselt number is computed on the heated wall of the Cu-water nanofluid-filled 3D cubic cavity, considering different nanoparticle volume fractions (\(\phi\)) and Rayleigh numbers (\(Ra\)). The table reveals that the maximum relative difference in Nusselt number values compared to Sannad et al. is 2.56\%, while the maximum relative difference compared to Ravnik et al. is just 1.29\%, across all parameters considered. It is also noteworthy that the maximum difference (2.56\%) with Sannad et al. occurs at $\phi=0.0$ and $Ra=10^5$. For these same parameters, the maximum difference with Ravnik et al. \cite{Ravnik_2010} is only 0.93\%. Hence, Table \ref{Average_Nusselt_Number_Comparison} clearly demonstrates that the results obtained from our code and the HOSC scheme are in good agreement with the existing results \cite{Ravnik_2010, Sannad_2020} on 3D nanofluid natural convection. Figure \ref{fig:3d_isotherm_comparison} provides a qualitative comparison of isotherm contours for various nanoparticle volume fractions ($\phi$) at $Ra=10^3$. The results show excellent agreement with the reference result \cite{Ravnik_2010}. Since our study involves the incorporation of conducting fins, we also compared our results (in Figure \ref{fig:fin_comparison}) with those of Ramón et al. \cite{Ramón_2007}, who analyzed a single conducting fin on the heated wall of a 3D cavity. The comparison of the temperature and velocity vector field demonstrates excellent agreement with their findings, confirming that our scheme accurately captures the flow phenomena even with the inclusion of a conducting fin. The thorough validation of the code and HOSC scheme, both quantitatively and qualitatively, confirms its effectiveness in simulating the 3D  natural convection of nanofluid.

\section{Results and Discussion}
\label{sec:Results and Discussion}
In this section, we explore and analyze the results of our research. Figure \ref{fig:Sche_diag} presents the schematic diagram of our research problem. The left wall is maintained at a non-dimensional temperature of 1, while the right wall is kept cold at $\theta =0$. The remaining four walls are insulated and considered adiabatic. No-slip boundary conditions are imposed on the velocities at each wall. Here, we examine two cases by using the HOSC scheme: the first involves the natural convection of Cu-water nanofluid within a 3D cubic cavity, and the second incorporates two conducting fins, each with a dimensionless width of 0.1 and length of 0.5, attached to the heated wall of the Cu-water nanofluid filled cubic cavity, as illustrated in Figure \ref{fig:Sche_diag}(b). We also include the gravitational force operating in the $y$ direction.
%%%%%%%%%%%%%%%%%%%%%%%%%%%%%%%%%%%%%%%%%%%%%%%%%%%%%%%%%%%%%%%%%%%%%%%%%%%%%%%%%%%%%%%%%%%%%
\subsection{Study of Flow Patterns and Temperature Contours}
\subsubsection{Case 1}

\begin{figure}[htbp]
 \centering
 \vspace*{5pt}%
 \hspace*{\fill}% 
\begin{subfigure}{0.50\textwidth}     % start subfigure 1
    %\belowcaptionskip=8pt
    \centering
    \includegraphics[width=\textwidth]{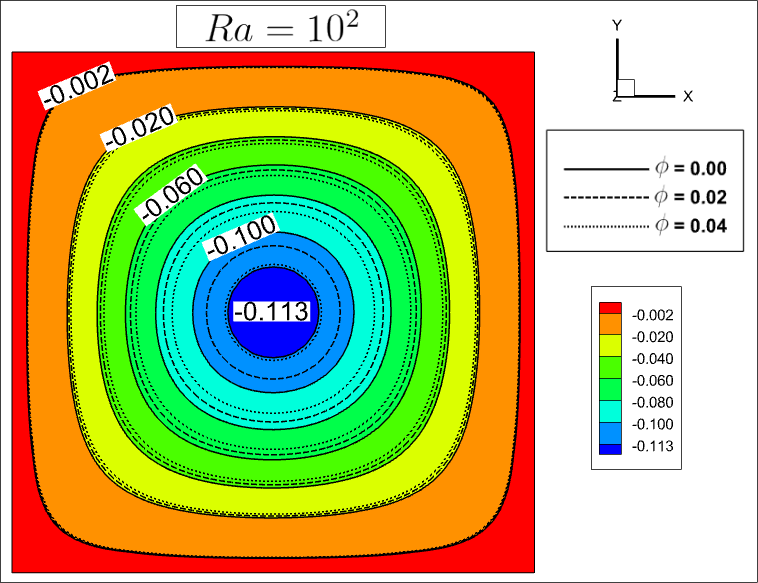}%
    \captionsetup{skip=5pt}%
    \caption{(a)}
    \label{fig:P1_Streamlines_Ra_10^2}
  \end{subfigure}%   
         % end subfigure 1
         % empty line absolutely necessary!
 \begin{subfigure}{0.50\textwidth}        % start subfigure 2
    %\belowcaptionskip=8pt
   \centering
    \includegraphics[width=\textwidth]{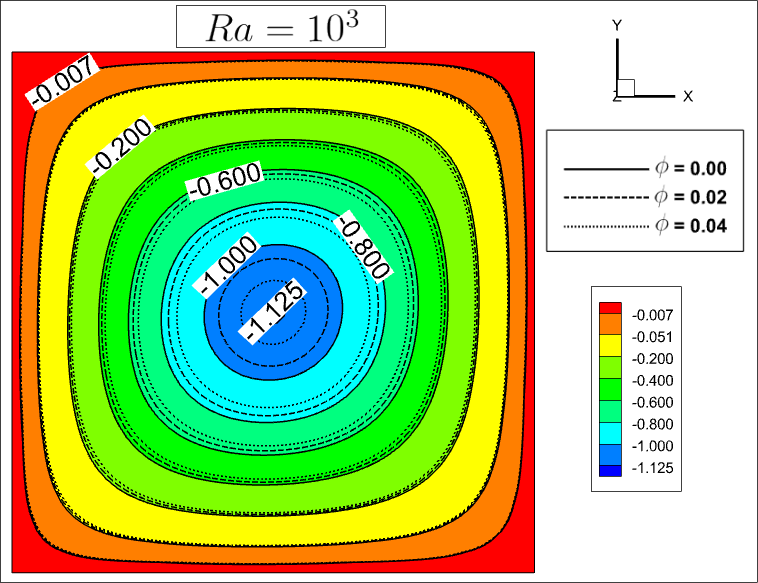}%
    \captionsetup{skip=5pt}%
    \caption{(b)}
    \label{fig:P1_Streamlines_Ra_10^3}
  \end{subfigure}%          % end subfigure 2  
  \hspace*{\fill}%          % empty line absolutely necessary!

  \vspace*{8pt}%
  \hspace*{\fill}%  
  \begin{subfigure}{0.50\textwidth}     % start subfigure 1
    %\belowcaptionskip=8pt
    \centering
    \includegraphics[width=\textwidth]{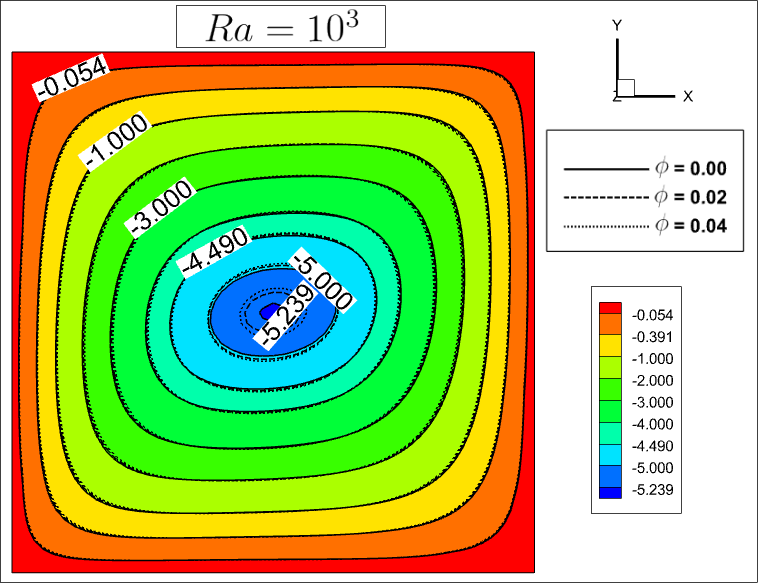}%
    \captionsetup{skip=5pt}%
    \caption{(c)}
    \label{fig:P1_Streamlines_Ra_10^4}
  \end{subfigure}%  
  \begin{subfigure}{0.50\textwidth}     % start subfigure 1
    %\belowcaptionskip=8pt
    \centering
    \includegraphics[width=\textwidth]{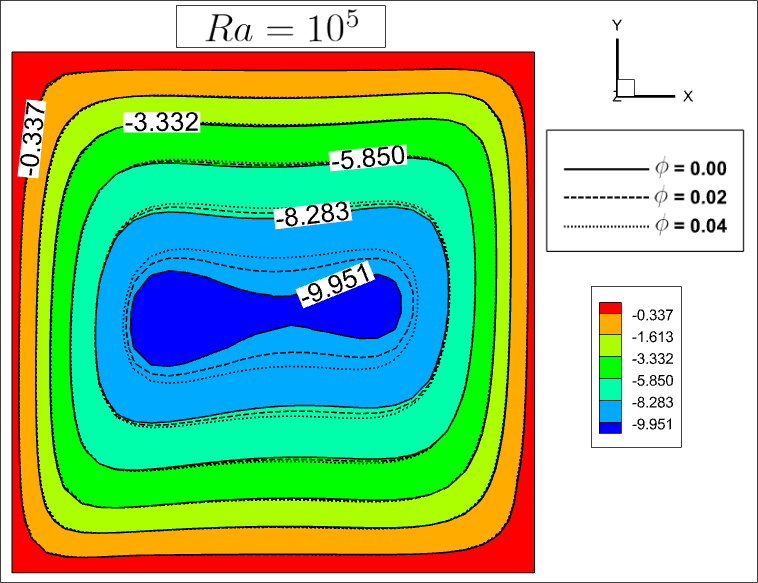}%
    \captionsetup{skip=5pt}%
    \caption{(d)}
    \label{fig:P1_Streamlines_Ra_10_5}
  \end{subfigure}%  
  \caption{Case 1: Streamlines of fluid at the symmetric ($z=0.5$) plane for cubic cavity flow under various $\phi$ and $Ra$ values: (a) $Ra = 10^2$, (b) $Ra = 10^3$, (c) $Ra = 10^4$, (d) $Ra = 10^5$}
  \label{fig:Streamlines_Contours_2d}
\end{figure}

%%%%%%%%%%%%%%%%%%%%%%%%%%%%%%%%%%%%%%%%%%%%%%%%%%%%%

\begin{figure}[htbp]
 \centering
 \vspace*{5pt}%
 \hspace*{\fill}% 
\begin{subfigure}{0.50\textwidth}     % start subfigure 1
    %\belowcaptionskip=8pt
    \centering
    \includegraphics[width=\textwidth]{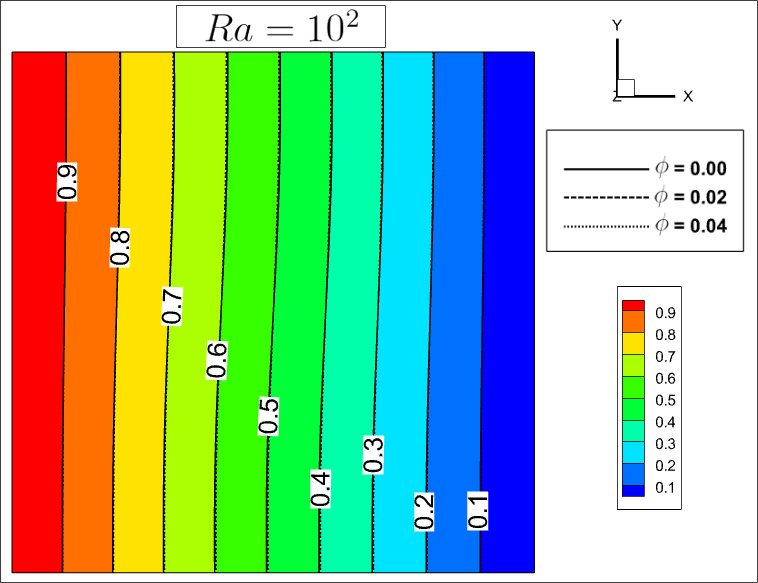}%
    \captionsetup{skip=5pt}%
    \caption{(a)}
    \label{fig:P1_Isotherm_Ra_10^2}
  \end{subfigure}%   
         % end subfigure 1
         % empty line absolutely necessary!
 \begin{subfigure}{0.50\textwidth}        % start subfigure 2
    %\belowcaptionskip=8pt
   \centering
    \includegraphics[width=\textwidth]{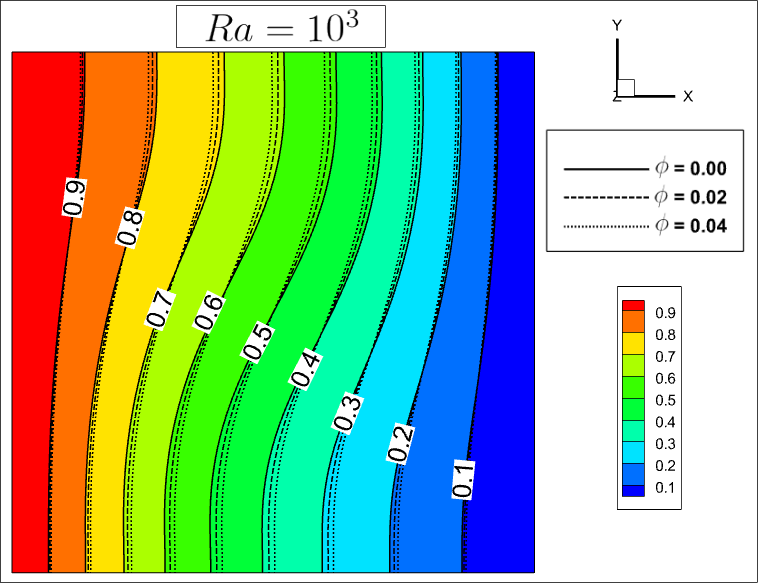}%
    \captionsetup{skip=5pt}%
    \caption{(b)}
    \label{fig:P1_Isotherm_Ra_10^3}
  \end{subfigure}%          % end subfigure 2  
  \hspace*{\fill}%          % empty line absolutely necessary!

  \vspace*{8pt}%
  \hspace*{\fill}%  
  \begin{subfigure}{0.50\textwidth}     % start subfigure 1
    %\belowcaptionskip=8pt
    \centering
    \includegraphics[width=\textwidth]{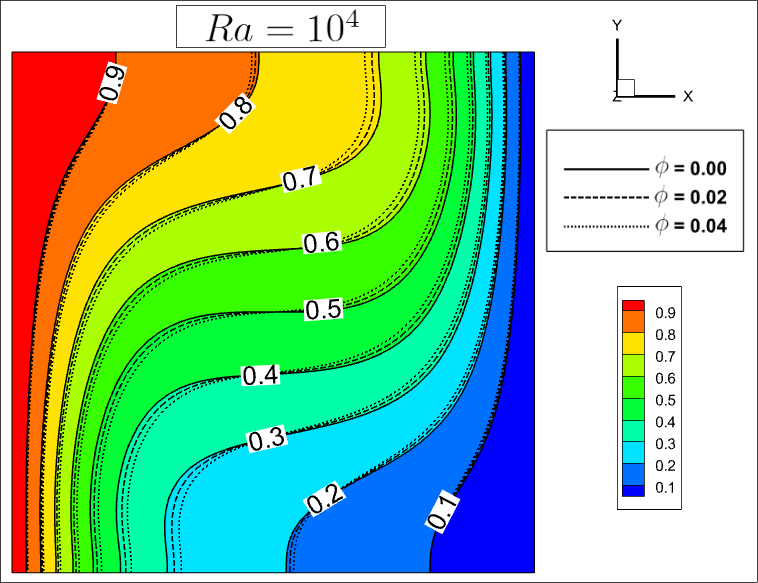}%
    \captionsetup{skip=5pt}%
    \caption{(c)}
    \label{fig:P1_Isotherm_Ra_10^4}
  \end{subfigure}%  
  \begin{subfigure}{0.50\textwidth}     % start subfigure 1
    %\belowcaptionskip=8pt
    \centering
    \includegraphics[width=\textwidth]{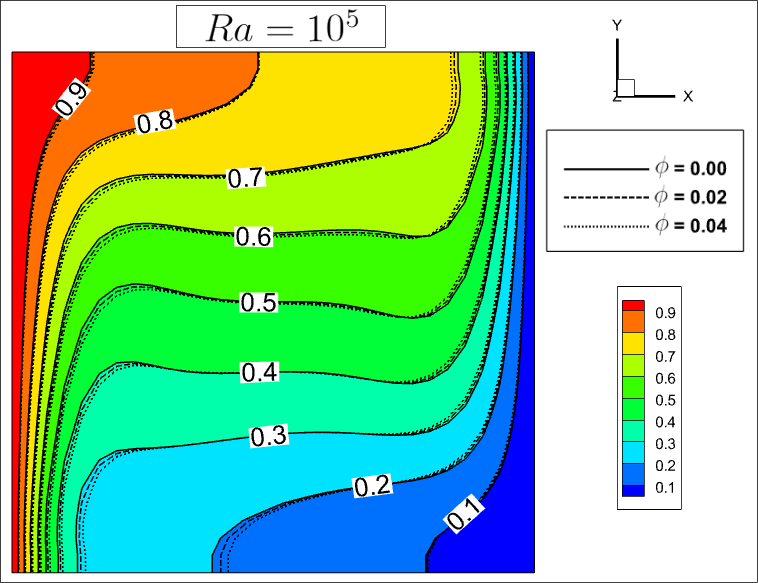}%
    \captionsetup{skip=5pt}%
    \caption{(d)}
    \label{fig:P1_Isotherm_Ra_10^5}
  \end{subfigure}%  
  \caption{Case 1: Isotherm contours at the symmetric plane ($z = 0.5$) under various $\phi$ and $Ra$ values: (a) $Ra = 10^2$, (b) $Ra = 10^3$, (c) $Ra = 10^4$, (d) $Ra = 10^5$}
  \label{fig:Case_1_Isotherm_Contours_2d}
\end{figure}

%%%%%%%%%%%%%%%%%%%%%%%%%%%%%%%%%%%%%%%%%%%%%%%%%%%%%

\begin{figure}[htbp]
 \centering
 \vspace*{5pt}%
 \hspace*{\fill}% 
\begin{subfigure}{0.50\textwidth}     % start subfigure 1
    %\belowcaptionskip=8pt
    \centering
    \includegraphics[width=\textwidth]{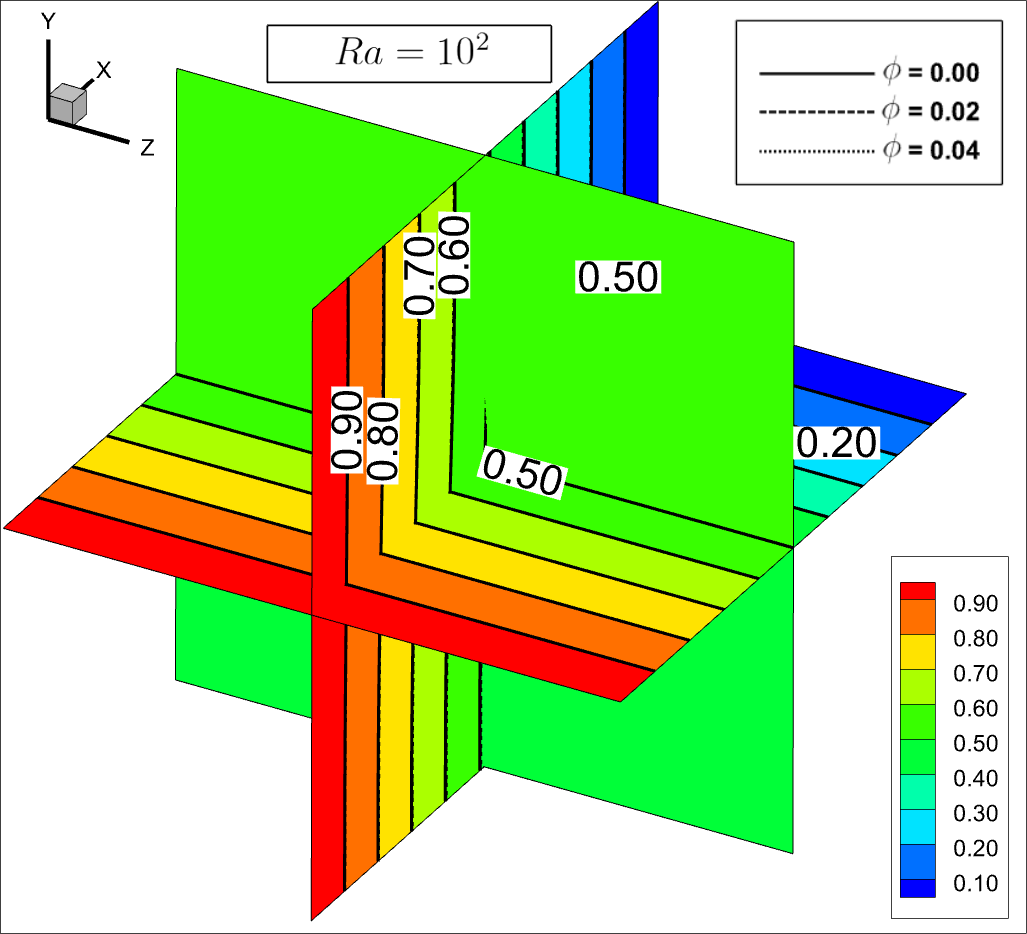}%
    \captionsetup{skip=5pt}%
    \caption{(a)}
    \label{fig:3D_P1_Isotherm_Ra_10^2}
  \end{subfigure}%   
         % end subfigure 1
         % empty line absolutely necessary!
 \begin{subfigure}{0.50\textwidth}        % start subfigure 2
    %\belowcaptionskip=8pt
   \centering
    \includegraphics[width=\textwidth]{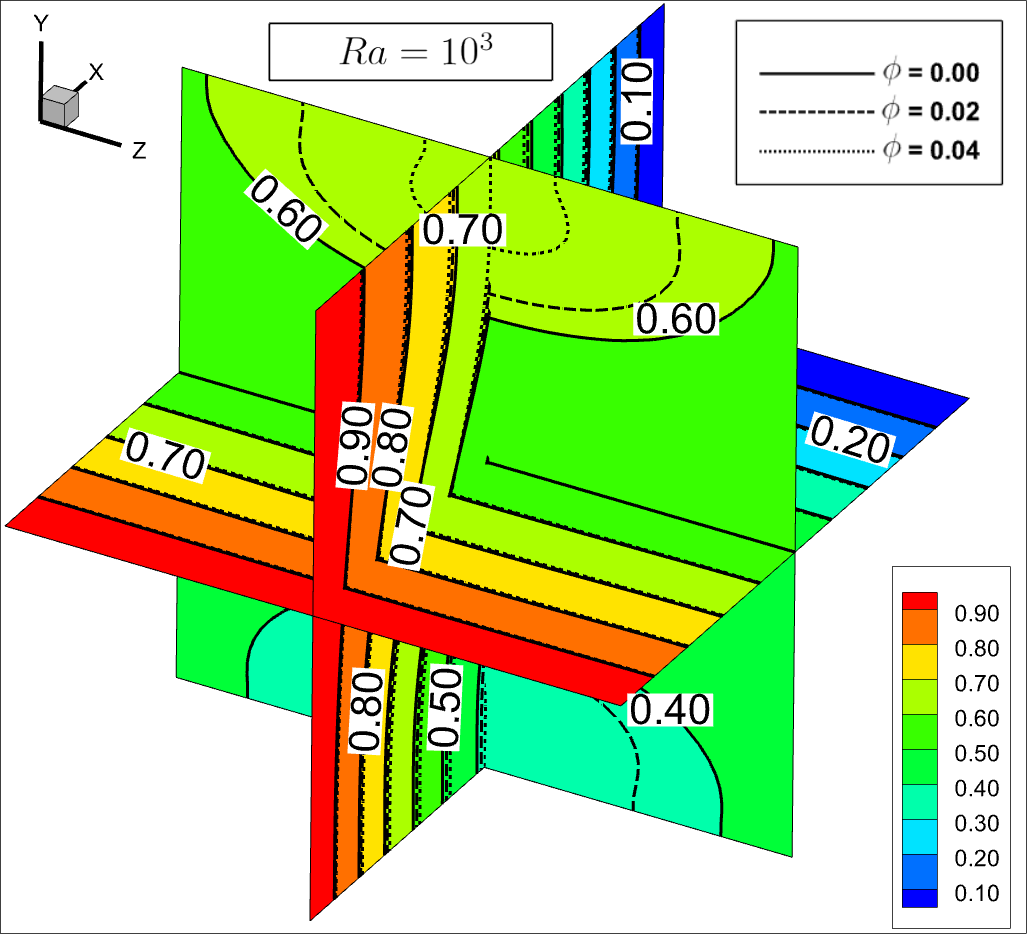}%
    \captionsetup{skip=5pt}%
    \caption{(b)}
    \label{fig:3D_P1_Isotherm_Ra_10^3}
  \end{subfigure}%          % end subfigure 2  
  \hspace*{\fill}%          % empty line absolutely necessary!

  \vspace*{8pt}%
  \hspace*{\fill}%  
  \begin{subfigure}{0.50\textwidth}     % start subfigure 1
    %\belowcaptionskip=8pt
    \centering
    \includegraphics[width=\textwidth]{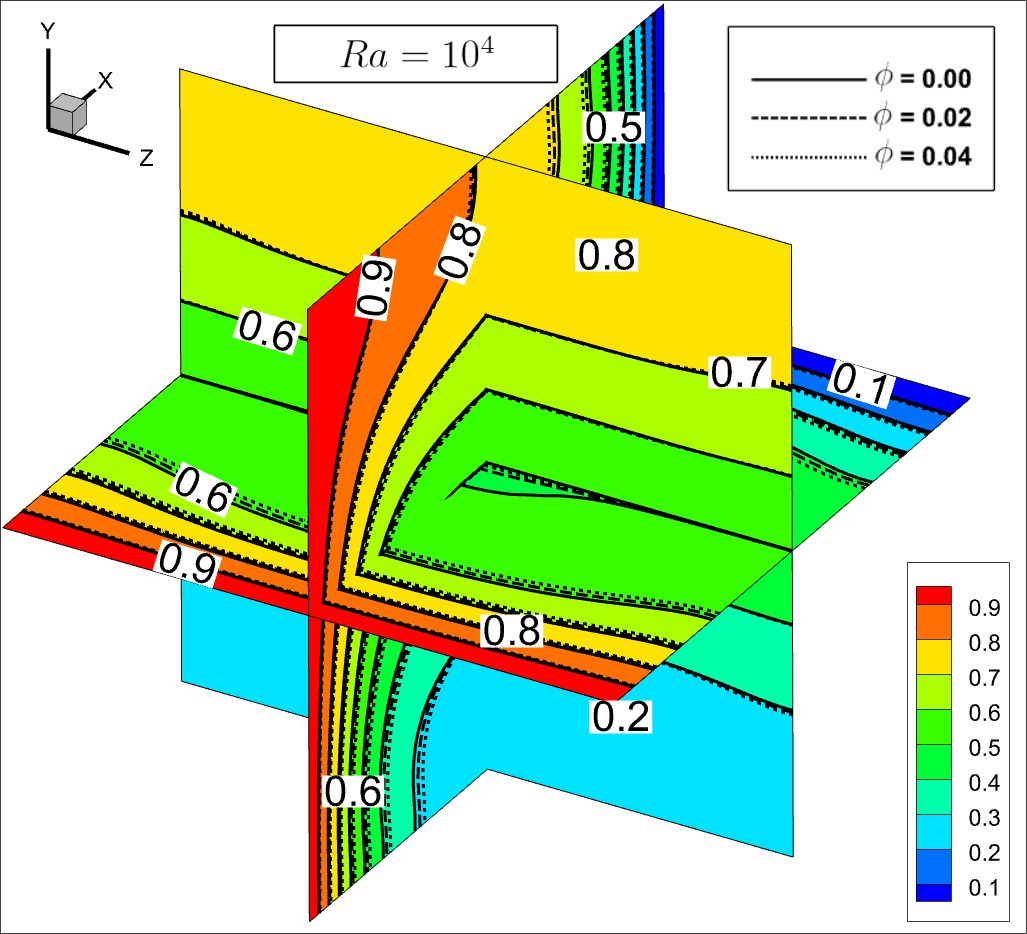}%
    \captionsetup{skip=5pt}%
    \caption{(c)}
    \label{fig:3D_P1_Isotherm_Ra_10^4}
  \end{subfigure}%  
  \begin{subfigure}{0.50\textwidth}     % start subfigure 1
    %\belowcaptionskip=8pt
    \centering
    \includegraphics[width=\textwidth]{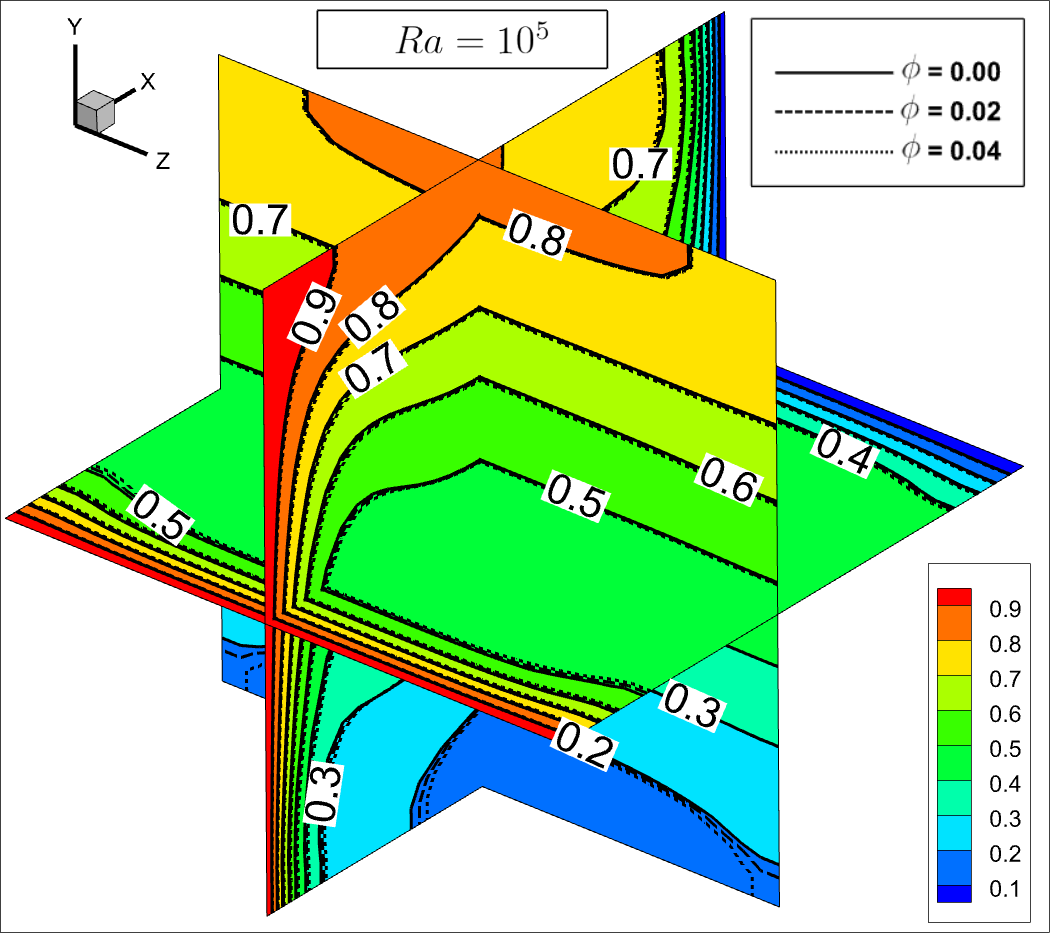}%
    \captionsetup{skip=5pt}%
    \caption{(d)}
    \label{fig:3D_P1_Isotherm_Ra_10^5}
  \end{subfigure}%  
  \caption{Case 1: Isotherm contours on three different symmetric planes ($y=0.5,z=0.5$ and $x=0.5$) for $0.00 \leq \phi \leq 0.04$ and $10^2\leq Ra\leq10^5$}
  \label{fig:Case_1_Isotherm_Contours_3d}
\end{figure}
Here, We explore the natural convection of nanofluid (Cu-water) inside a 3D cubic cavity by examining the streamlines and temperature distribution across a range of Rayleigh number ($10^2 \leq Ra \leq 10^5$) and nanoparticle volume fractions ($0.00 \leq \phi \leq 0.04$) for Case 1. Figure \ref{fig:Streamlines_Contours_2d} demonstrates the effect of varying $Ra$ and $\phi$ on the streamlines within the symmetric plane at $z=0.5$. The figure compares the flow fields for pure fluid (solid line), nanofluid with $\phi=0.02$ (dashed line), and nanofluid with $\phi=0.04$ (dotted line). At lower Rayleigh numbers ($Ra=10^2, 10^3$), the circular pattern of the streamlines is observed, indicating the dominance of viscous forces over buoyancy. When the Rayleigh number reaches $Ra=10^4$, the streamline patterns transition from circular to ellipsoidal shape. Further increasing $Ra$ to $10^5$ results in more pronounced streamlines, with the appearance of multiple vortices. The central primary elliptical vortex in the cavity splits into two distinct vortices. This division emphasizes the increasing dominance of buoyancy forces, which significantly alter the flow dynamics, leading to more complex and intensified convective motion within the cavity. In all cases, the streamfunction values remain negative, signifying a clockwise rotational flow. The introduction of nanoparticles alters the cell's diameter, with a noticeable increase in diameter as $\phi$ increases, particularly near the middle part of the cavity for each $Ra$ value. The difference in vortex diameters becomes more noticeable at low Rayleigh numbers. The reason is that at lower $Ra$, where buoyancy forces are weaker, the addition of nanoparticles significantly alters the flow structure, leading to a greater difference in the diameter of the streamlines cells between nanofluid and pure fluid. However, as $Ra$ increases and buoyancy forces grow stronger, this difference in streamlines cell size diminishes, suggesting that the impact of nanoparticles on the flow structure diminishes at higher $Ra$. Furthermore, the maximum magnitude of the streamfunction increases with higher $Ra$ for both nanofluid and pure fluid, indicating that the flow intensity and complexity are enhanced as buoyancy forces become more dominant.\\
Figures \ref{fig:Case_1_Isotherm_Contours_2d} and \ref{fig:Case_1_Isotherm_Contours_3d} prominently depict the isothermal contours on the symmetric planes, highlighting the effects of different $Ra$ and $\phi$ values.
For comparative purposes, the isotherms for pure fluid (water) and nanofluid (Cu-water) with volume fractions $\phi=0.02$ and $0.04$ are displayed in the same figure.
The isotherms on the symmetric plane at $z=0.5$ are presented separately in Figure \ref{fig:Case_1_Isotherm_Contours_2d} to enhance clarity and provide a more detailed visualization of the temperature distribution, particularly because this plane experiences significant heat transfer from the heated wall to the cooler wall. Isotherm patterns at $z = 0.5$ display diagonal symmetry for all values of $Ra$ and $\phi$. For low $Ra$ values ($10^2$), the isotherm lines remain horizontal and parallel with thermally active vertical walls, indicating a low intensity of natural convection. The impact of volume fraction on the thermal distribution is negligible at this $Ra$ value. As the $Ra$ values increase to $10^3$ and $10^4$, the thermal boundary layers thin out, and the thermal contour lines shift from straight to curved. This change occurs because a higher $Ra$ corresponds to enhanced buoyancy forces, making heat transfer more convection-dominant for both pure fluids and nanofluid. For $Ra=10^5$, the isotherms near the center of the cavity become predominantly horizontal, while they maintain a vertical orientation only within the narrow boundary layers adjacent to the active walls. These isotherms also demonstrate that for $Ra \geq 10^3$, thicker thermal boundary layers occur when the volume percentage of nanoparticles increases. This effect is due to the enhanced heat conductivity of the nanofluid, which lowers the temperature difference near the active walls and enhances heat diffusion. The response of the thickness of the thermal boundary layer to nanoparticle volume percentage demonstrates the importance of improved thermal conductivity in nanofluid. Additionally, an analysis of the other two symmetry planes ($x = 0.5$ $\&$ $y = 0.5$) reveals that as $Ra$ increases, the temperature rises as it approaches the center of the $x = 0.5$ plane, while it decreases towards the center of the $y = 0.5$ plane for both pure fluid and nanofluid. The influence of $\phi$ on the $y=0.5$ plane is similar to the effects observed on the $z=0.5$ plane. Also, as $Ra$ increases, the isotherm lines on the $x=0.5$ plane become more distinct and spread out. At lower $Ra$, buoyancy forces are weak relative to viscous forces, leading to less pronounced isotherm patterns. In contrast, higher $Ra$ values indicate stronger buoyancy forces compared to viscous forces, resulting in more pronounced and widely spaced isotherms.

%%%%%%%%%%%%%%%%%%%%%%%%%%%%%%%%%%%%%%%%%%%%%%%%%%%%%
\subsubsection{Case 2}
\begin{figure}[htbp]
 \centering
 \vspace*{5pt}%
 \hspace*{\fill}% 
\begin{subfigure}{0.50\textwidth}     % start subfigure 1
    %\belowcaptionskip=8pt
    \centering
    \includegraphics[width=\textwidth]{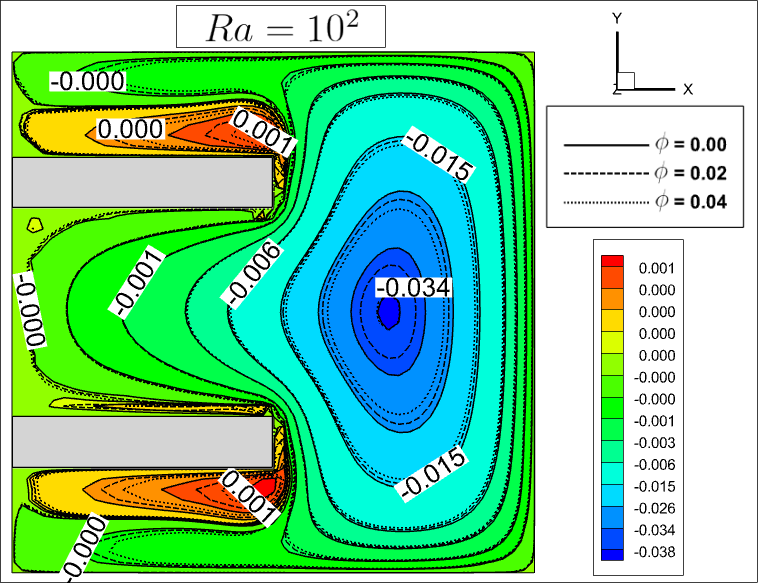}%
    \captionsetup{skip=5pt}%
    \caption{(a)}
    \label{fig:P2_Streamlines_Ra_10^2}
  \end{subfigure}%   
         % end subfigure 1
         % empty line absolutely necessary!
 \begin{subfigure}{0.50\textwidth}        % start subfigure 2
    %\belowcaptionskip=8pt
   \centering
    \includegraphics[width=\textwidth]{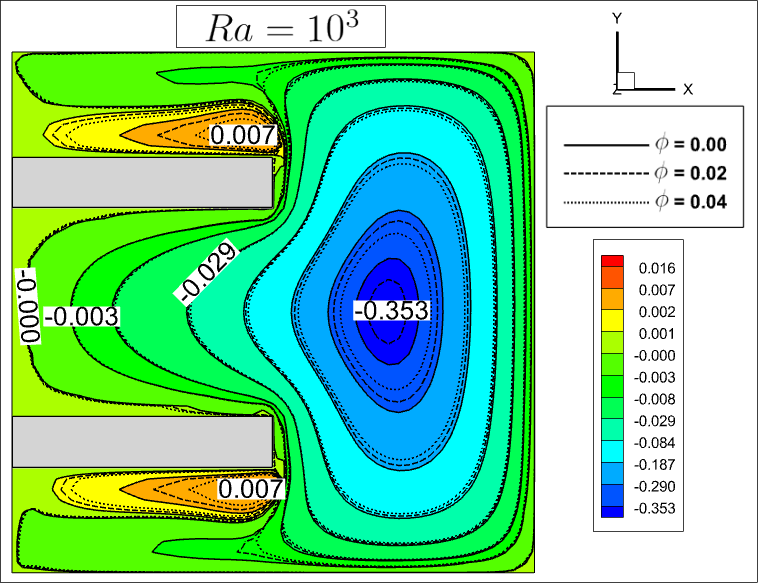}%
    \captionsetup{skip=5pt}%
    \caption{(b)}
    \label{fig:P2_Streamlines_Ra_10^3}
  \end{subfigure}%          % end subfigure 2  
  \hspace*{\fill}%          % empty line absolutely necessary!

  \vspace*{8pt}%
  \hspace*{\fill}%  
  \begin{subfigure}{0.50\textwidth}     % start subfigure 1
    %\belowcaptionskip=8pt
    \centering
    \includegraphics[width=\textwidth]{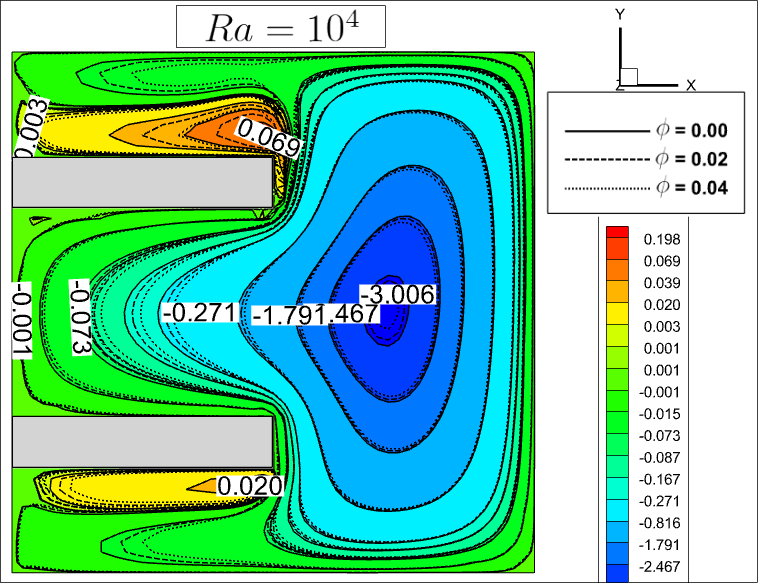}%
    \captionsetup{skip=5pt}%
    \caption{(c)}
    \label{fig:P2_Streamlines_Ra_10^4}
  \end{subfigure}%  
  \begin{subfigure}{0.50\textwidth}     % start subfigure 1
    %\belowcaptionskip=8pt
    \centering
    \includegraphics[width=\textwidth]{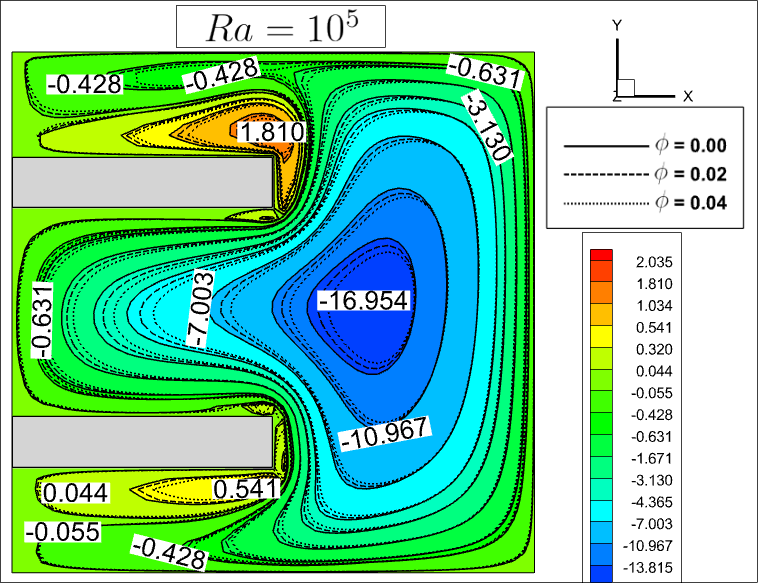}%
    \captionsetup{skip=5pt}%
    \caption{(d)}
    \label{fig:P2_Streamlines_Ra_10^5}
  \end{subfigure}%  
  \caption{Case 2: Streamlines of fluid at the symmetric ($z=0.5$) plane for cubic cavity flow under various $\phi$ and $Ra$ values: (a) $Ra = 10^2$, (b) $Ra = 10^3$, (c) $Ra = 10^4$, (d) $Ra = 10^5$}
  \label{fig:Case_2_Streamlines_Contours_2d}
\end{figure}

\begin{figure}[htbp]
 \centering
 \vspace*{5pt}%
 \hspace*{\fill}% 
\begin{subfigure}{0.50\textwidth}     % start subfigure 1
    %\belowcaptionskip=8pt
    \centering
    \includegraphics[width=\textwidth]{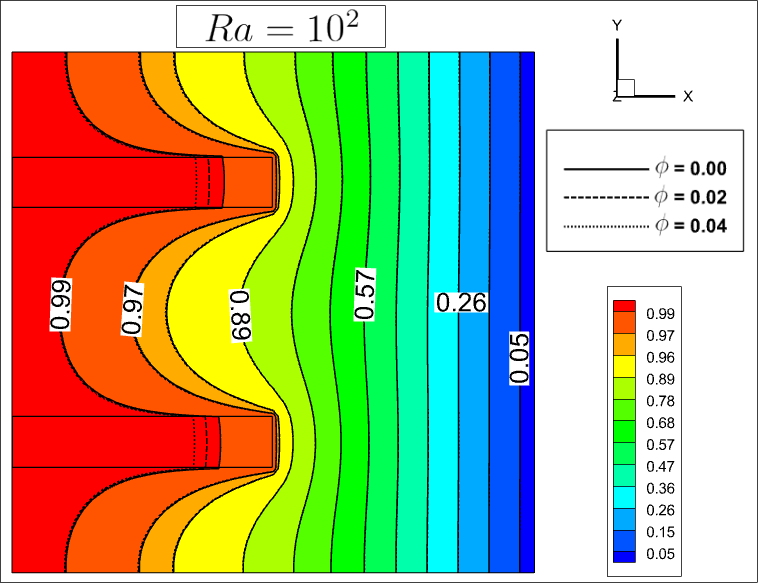}%
    \captionsetup{skip=5pt}%
    \caption{(a)}
    \label{fig:P2_Isotherm_Ra_10^2}
  \end{subfigure}%   
         % end subfigure 1
         % empty line absolutely necessary!
 \begin{subfigure}{0.50\textwidth}        % start subfigure 2
    %\belowcaptionskip=8pt
   \centering
    \includegraphics[width=\textwidth]{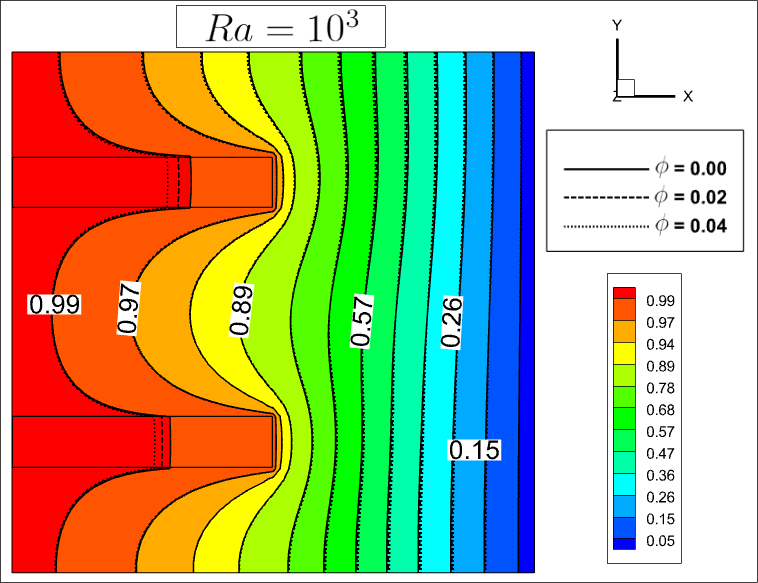}%
    \captionsetup{skip=5pt}%
    \caption{(b)}
    \label{fig:P2_Isotherm_Ra_10^3}
  \end{subfigure}%          % end subfigure 2  
  \hspace*{\fill}%          % empty line absolutely necessary!

  \vspace*{8pt}%
  \hspace*{\fill}%  
  \begin{subfigure}{0.50\textwidth}     % start subfigure 1
    %\belowcaptionskip=8pt
    \centering
    \includegraphics[width=\textwidth]{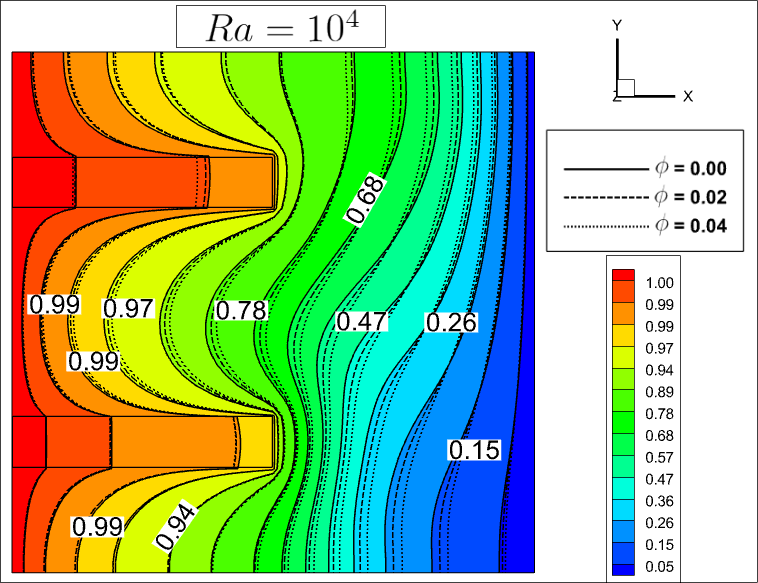}%
    \captionsetup{skip=5pt}%
    \caption{(c)}
    \label{fig:P2_Isotherm_Ra_10^4}
  \end{subfigure}%  
  \begin{subfigure}{0.50\textwidth}     % start subfigure 1
    %\belowcaptionskip=8pt
    \centering
    \includegraphics[width=\textwidth]{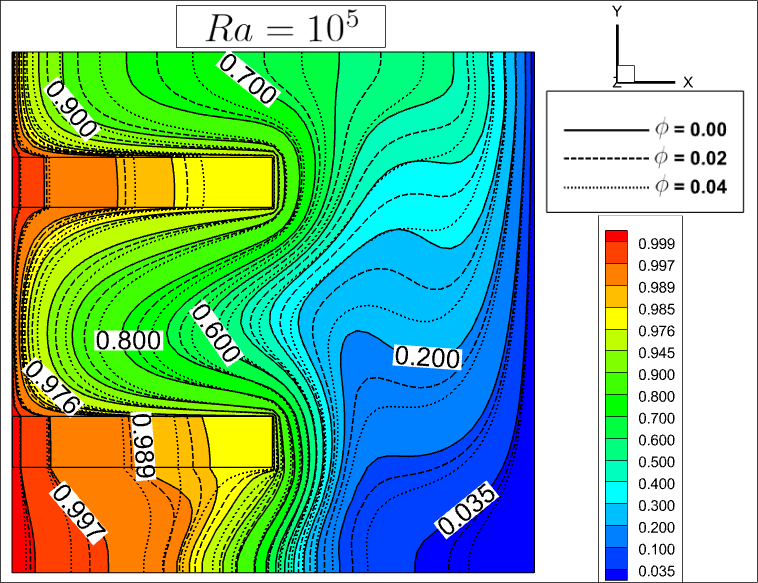}%
    \captionsetup{skip=5pt}%
    \caption{(d)}
    \label{fig:P2_Isotherm_Ra_10^5}
  \end{subfigure}%  
  \caption{Case 2: Isotherm contours at the symmetric plane ($z = 0.5$) under various $\phi$ and $Ra$ values: (a) $Ra = 10^2$, (b) $Ra = 10^3$, (c) $Ra = 10^4$, (d) $Ra = 10^5$}
  \label{fig:Isotherm_Contours_2d}
\end{figure}

\begin{figure}[htbp]
 \centering
 \vspace*{5pt}%
 \hspace*{\fill}% 
\begin{subfigure}{0.50\textwidth}     % start subfigure 1
    %\belowcaptionskip=8pt
    \centering
    \includegraphics[width=\textwidth]{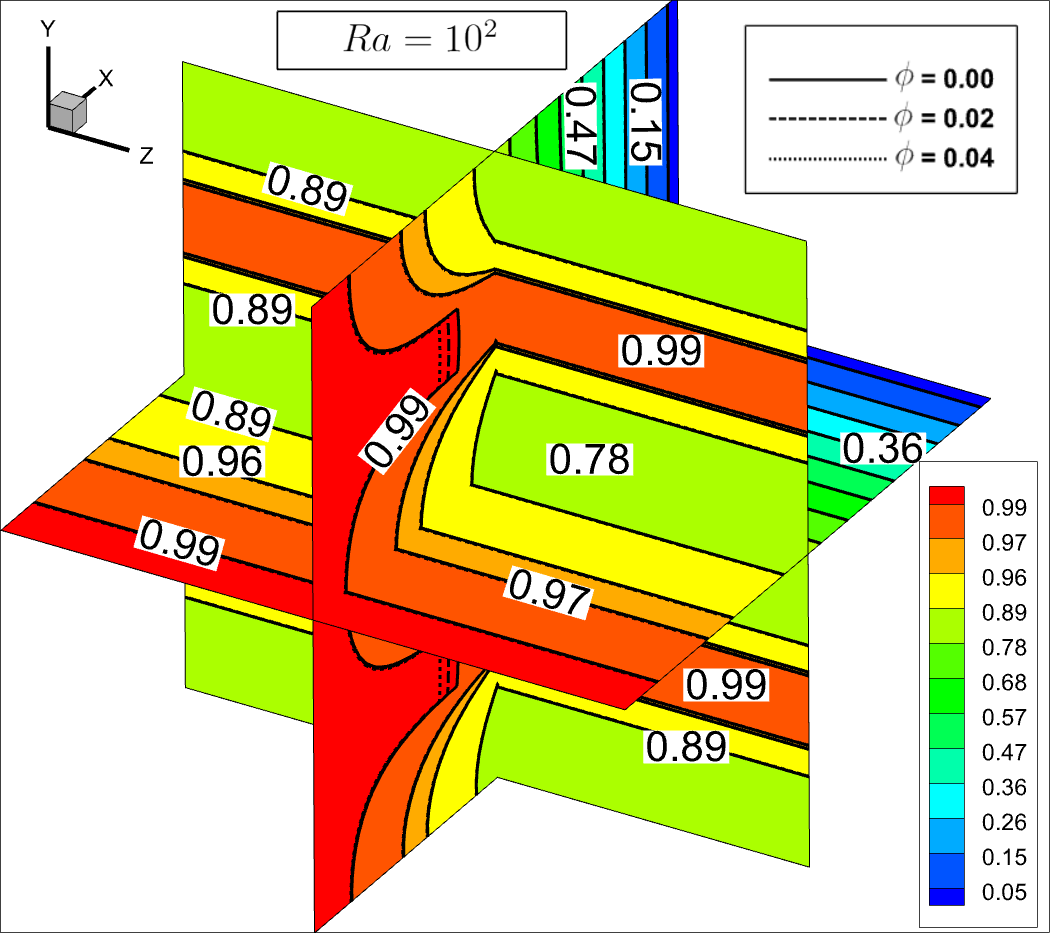}%
    \captionsetup{skip=5pt}%
    \caption{(a)}
    \label{fig:3D_P2_Isotherm_Ra_10^2}
  \end{subfigure}%   
         % end subfigure 1
         % empty line absolutely necessary!
 \begin{subfigure}{0.50\textwidth}        % start subfigure 2
    %\belowcaptionskip=8pt
   \centering
    \includegraphics[width=\textwidth]{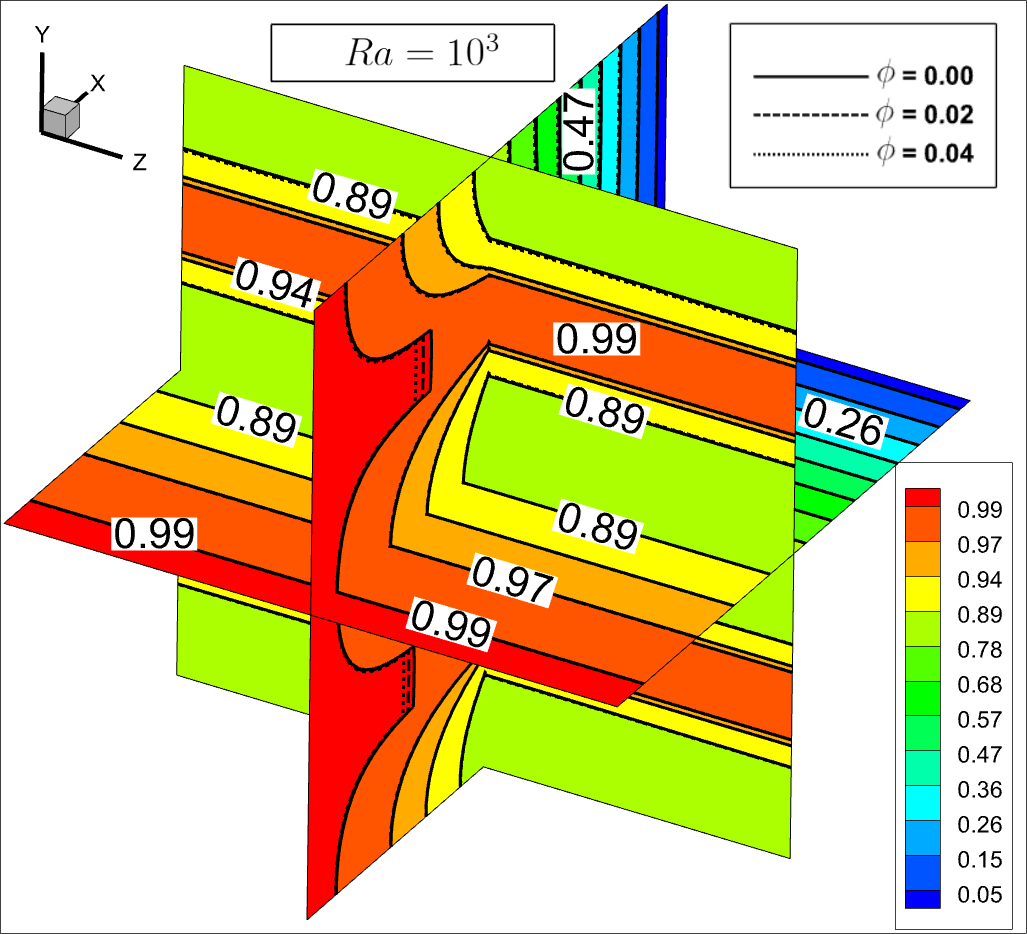}%
    \captionsetup{skip=5pt}%
    \caption{(b)}
    \label{fig:3D_P2_Isotherm_Ra_10^3}
  \end{subfigure}%          % end subfigure 2  
  \hspace*{\fill}%          % empty line absolutely necessary!

  \vspace*{8pt}%
  \hspace*{\fill}%  
  \begin{subfigure}{0.50\textwidth}     % start subfigure 1
    %\belowcaptionskip=8pt
    \centering
    \includegraphics[width=\textwidth]{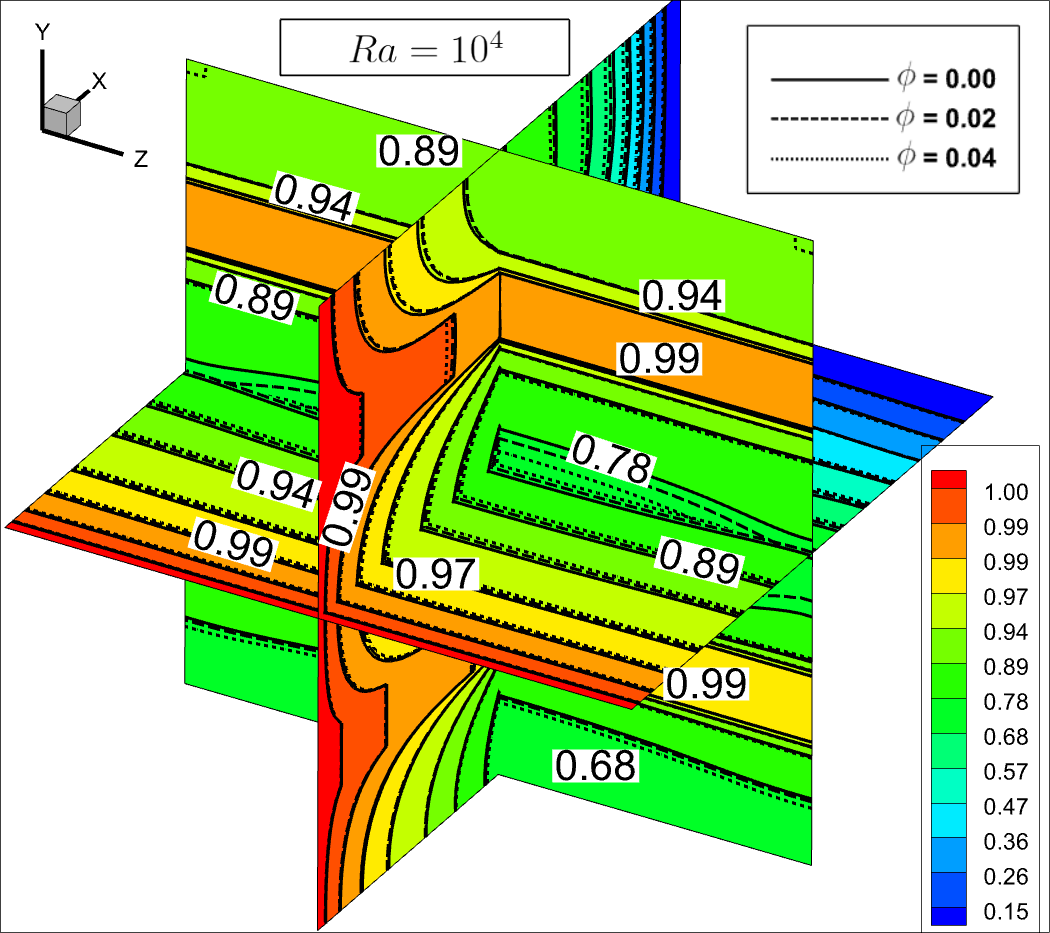}%
    \captionsetup{skip=5pt}%
    \caption{(c)}
    \label{fig:3D_P2_Isotherm_Ra_10^4}
  \end{subfigure}%  
  \begin{subfigure}{0.50\textwidth}     % start subfigure 1
    %\belowcaptionskip=8pt
    \centering
    \includegraphics[width=\textwidth]{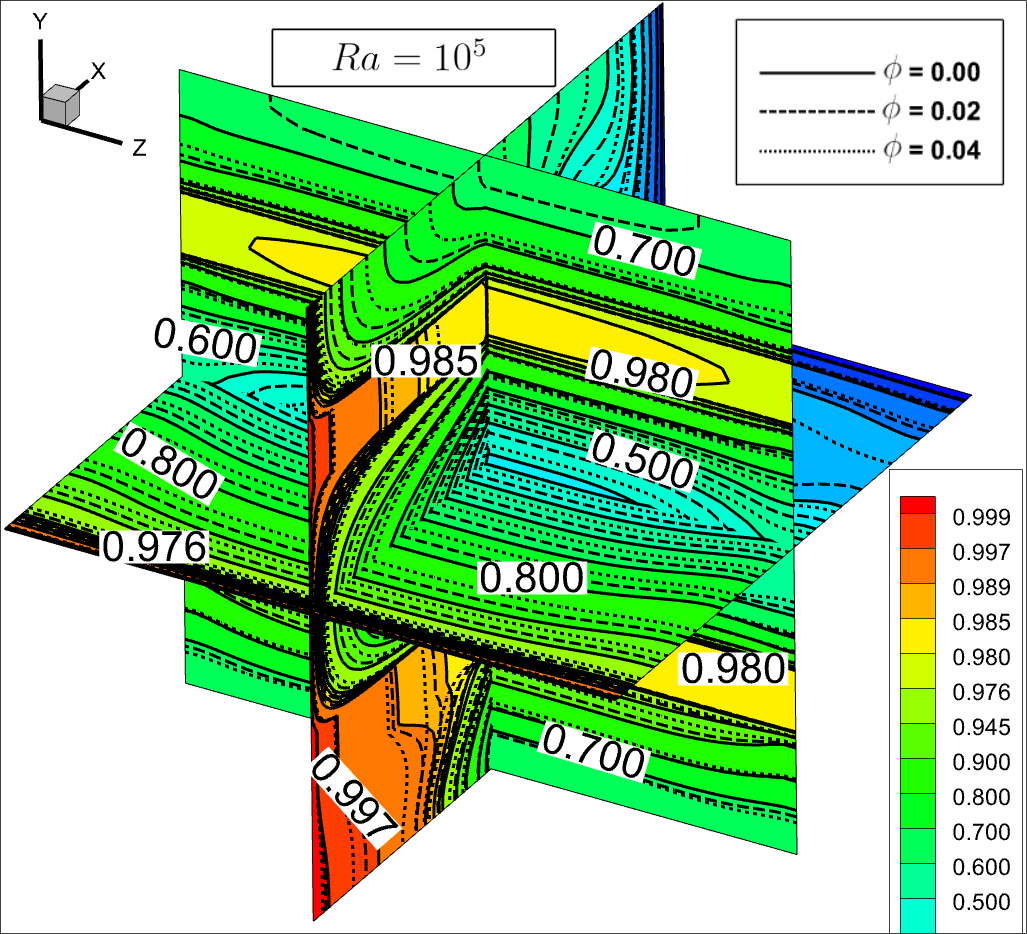}%
    \captionsetup{skip=5pt}%
    \caption{(d)}
    \label{fig:3D_P2_Isotherm_Ra_10^5}
  \end{subfigure}%  
  \caption{Case 2: Isotherm contours on three different symmetric planes ($y=0.5,z=0.5$ and $x=0.5$) for $0.00 \leq \phi \leq 0.04$ and $10^2\leq Ra\leq10^5$ }
  \label{fig:Case_2_Isotherm_Contours_3d}
\end{figure}

In Case 2, two solid conducting fins made of aluminum are placed on the heated wall of the cubic cavity. These fins have a dimensionless length of 0.5 and a width of 0.1, as depicted in the schematic diagram (Figure \ref{fig:Sche_diag}(b)). The cavity is filled with the same nanofluid (Cu-water), Similar to Case 1. Both fins are positioned 0.2 distance (dimensionless) away from the center of the left wall. In this section, we examine how nanoparticle concentration, Rayleigh number, and the presence of conducting fins influence the flow patterns and temperature contours. Figure \ref{fig:Case_2_Streamlines_Contours_2d} presents the streamline patterns on the symmetric plane ($z=0.5$) for different Rayleigh numbers ($Ra$) and nanoparticle volume fractions ($\phi$). For \( Ra = 10^2 \), a primary vortex forms near the center of the cavity, accompanied by multiple secondary vortices near the conducting fins. The presence of these fins induces two symmetrical secondary vortices of equal size, strength, and shape-one above the upper fin and one below the lower fin. Additionally, a secondary vortex emerges near the top of the lower fin, but no corresponding vortex is observed near the bottom of the upper fin. The interaction of the main vortex with the solid conducting fins generates these secondary vortices. The streamfunction values for the secondary vortices are positive, indicating anticlockwise rotation, whereas the streamfunction values around the primary vortex are negative, signifying clockwise rotation of the flow. As the nanoparticle volume fraction (\(\phi\)) increases, the diameter of the streamline cells, particularly near the center of the primary vortex and between the fins, decreases. This reduction occurs because a higher nanoparticle concentration increases fluid viscosity, slowing down the flow. This behavior highlights the impact of nanoparticle concentration on the fluid dynamics within the cavity, a key finding in this study. The streamline patterns remain largely consistent for \(Ra = 10^3\) across all \(\phi\) values. However, there is a noticeable change in the magnitude of the streamfunction, indicating that while the flow dynamics are similar, the speed of the flow varies with $Ra$ and \(\phi\). Additionally, the small secondary vortex near the top of the lower fin tip gradually vanishes. At higher Rayleigh numbers (\(Ra = 10^4\) and \(Ra = 10^5\)), the shapes of the primary and secondary vortices undergo significant changes. The streamlines become denser between the two conducting fins, signifying increased fluid movement due to stronger buoyancy forces as \(Ra\) rises. The secondary vortex near the top of the lower conducting fin vanishes at $Ra=10^4$, but grows again at $Ra=10^5$. Across all \(Ra\) values, an increase in \(\phi\) consistently results in a reduction in the cell diameter of the streamlines. This is attributed to the increased nanoparticle concentration, which enhances the fluid's viscosity, slowing down the flow. Moreover, the maximum magnitude of the streamfunction for both primary and secondary vortices increases with \(Ra\). This phenomenon occurs because, as \(Ra\) increases, buoyancy forces become more dominant, leading to stronger and more intricate convection patterns. These findings emphasize the intricate relationship between nanoparticle concentration and Rayleigh number in influencing the flow dynamics within the cavity, underscoring the critical need to precisely adjust these parameters in applications involving nanofluid. 
\\
In Figures \ref{fig:Isotherm_Contours_2d} and \ref{fig:Case_2_Isotherm_Contours_3d}, we present the temperature distribution on the symmetric planes for Case 2. On the \(z=0.5\) plane (in Figure \ref{fig:Case_2_Isotherm_Contours_3d}), it is evident that for \(Ra=10^2\), the heat distributes uniformly from both the upper and lower sides of the heated wall, indicating a predominance of conduction over convection. The figure also compares isotherms for pure fluid and nanofluid, highlighting that the thermal distribution is almost similar between pure and naofluid, except in the conducting fins. Within the conducting fins, the isotherms are observed to be closer to the heated wall with higher nanoparticle volume fractions (\(\phi\)). This phenomenon is attributed to the improved thermal conductivity of the nanofluid. A similar isotherm pattern is noted for \(Ra=10^3\); however, the difference between the isotherms for various \(\phi\) values becomes less pronounced.   At \(Ra=10^4\), the dominance of buoyancy forces drives intense convective flows that markedly increase the heat transfer. The isotherms become tightly packed near the heated wall, reflecting steep temperature gradients and intensified heat transfer. Near the bottom of the heated wall, the thermal boundary layer forms with a pronounced gradient, resulting in closely spaced isotherms. As the fluid rises and cools, the gradient weakens, leading to wider spacing of isotherms, particularly towards the top of the cavity where convection effects are less intense. At \(Ra=10^5\), the isotherms exhibit a markedly curved pattern. Here, the spacing of isotherms near the top of the heated wall becomes narrower compared to the bottom, a shift from the previous lower $Ra$ values. This new phenomenon is observed for the first time. The increased buoyancy forces at this higher \(Ra\) induce more intense convection, accelerating the fluid's upward movement along the heated wall. The conducting fins on the heated wall significantly influence this behavior by redistributing heat and enhancing heat transfer through increased surface area and localized high heat flux regions. As the fluid rises more rapidly, it transfers heat more effectively to the surrounding fluid, leading to denser isotherms closer to the upper wall. Additionally, the influence of nanoparticle volume fraction (\(\phi\)) also reverses at this \(Ra\) value. The enhanced convection and heat transfer induced by the fins become more pronounced, illustrating the critical role of these fins in optimizing thermal performance in high Rayleigh number scenarios. Furthermore, observations of the other two symmetry planes (\( x = 0.5 \) $\&$ \( y = 0.5 \)) reveal that, as \( Ra \) increases, the temperature rises as it approaches the center of the \( x = 0.5 \) plane while decreasing as it approaches the center of the \( y = 0.5 \) plane, across all values of \( \phi \). This study of isotherms highlights how the interaction of buoyancy forces, conducting fins, and nanoparticle concentration can dramatically affect temperature distribution and heat transfer dynamics in a 3D cavity.
%%%%%%%%%%%%%%%%%%%%%%%%%%%%%%%%%%%%%%%%%%%%%%%%%%%%%%%%%%%%%%%%%%%%%%%%%%%%%%%%%%%%%%%%%%%%%

%%%%%%%%%%%%%%%%%%%%%%%%%%%%%%%%%%%%%%%%%%%%%%%%%%%%%%%%%%%%%%%%%%%%%%%%%%%%%%%%%%%%%%%%%%%%%

%%%%%%%%%%%%%%%%%%%%%%%%%%%%%%%%%%%%%%%%%%%%%%%%%%%%%

\subsubsection{Comparsion between Case 1 and Case 2}
The introduction of two aluminum conducting fins in Case 2 significantly alters the flow patterns and heat transfer characteristics compared to Case 1. In Case 1, where no fins are present, the streamline patterns exhibit a predominantly circular shape for \(Ra \leq 10^3\), featuring a single primary vortex. In comparison, Case 2, with the addition of conducting fins, displays a non-circular form of the primary vortex and multiple secondary vortices rotating in a clockwise direction—phenomena not observed in Case 1. At \(Ra=10^5\), the primary vortex in Case 1 splits into two distinct vortices, whereas in Case 2, a single primary vortex persists with secondary vortices still present. Regarding isotherms, both cases display relatively uniform distribution at \(Ra=10^2\) and \(10^3\). However, in Case 1, the isotherms remain nearly linear across the \(z=0.5\) plane, but in the scenario of Case 2, the conducting fins introduce localized alterations in isotherm patterns. Isotherms curved around the fins, displaying sharp temperature gradients near the fin surfaces because of the excellent thermal conductivity of the aluminum fins that spread heat from the heated wall into the fluid. In Case 1, the isotherms are close together near the bottom of the heated wall for \(Ra=10^5\). However, in Case 2, the presence of the conducting fin and high buoyancy effect causes a significant shift in temperature gradients from the bottom to the top of the heated wall at \(Ra=10^5\) for all values of $\phi$. This new phenomenon, observed for the first time, underscores the substantial impact of conducting fins on convection and heat transfer inside the cavity. The isotherms in Case 2 exhibit a large high-temperature region compared to Case 1 at each $Ra$ value, highlighting the fins' role in enhancing heat transfer by increasing the effective surface area in contact with the convective medium. 
It is also noteworthy that in Case 1, the variation of isotherms concerning $\phi$ (at $Ra=10^5$) is less pronounced compared to Case 2, indicating a more significant effect of $\phi$ on the temperature distribution in Case 2. This highlights the crucial role of $\phi$ in influencing the isotherm patterns and overall heat transfer.

\subsection{Analysis of Heat Transfer: Nusselt Number Study}
The Nusselt number ($Nu$) is a dimensionless quantity used in heat transfer to evaluate the effectiveness of convective heat transfer between a solid surface and the surrounding fluid. It quantifies how effectively heat is transferred from a solid surface to the fluid through convection. Local Nusselt number at heated wall $(x=0.0)$ for nanofluid-filled cavity can be defined as follows:
$$Nu_{\text {L }}(y, z)= -\frac{k_{nf}}{k_{bf}}\frac{\partial \theta(y, z)}{\partial x} $$ 
The local Nusselt number is specified in the same manner for Case 2, except at the fin base, where it is determined as \cite{Ramón_2007}:
$$Nu_{\text {L }}(y, z)= -\frac{k_{s}}{k_{bf}}\frac{\partial \theta(y, z)}{\partial x}$$ 
Total Nusselt number($Nu_{\text {T}}$) is defined as follows: 
$$N u_{\text {T}}=\int_0^1 N u_{\text {Avg}}(z) \mathrm{d} z .
$$
Here, $Nu_{\text{Avg}}$ denotes the average Nusselt number, determined by integrating the local Nusselt number over the relevant surface.\\

\begin{figure}[htbp]
 \centering
 \vspace*{0pt}%
 \hspace*{\fill}% 
\begin{subfigure}{1.0\textwidth}     % start subfigure 1
    %\belowcaptionskip=8pt
    \centering
    \includegraphics[width=\textwidth]{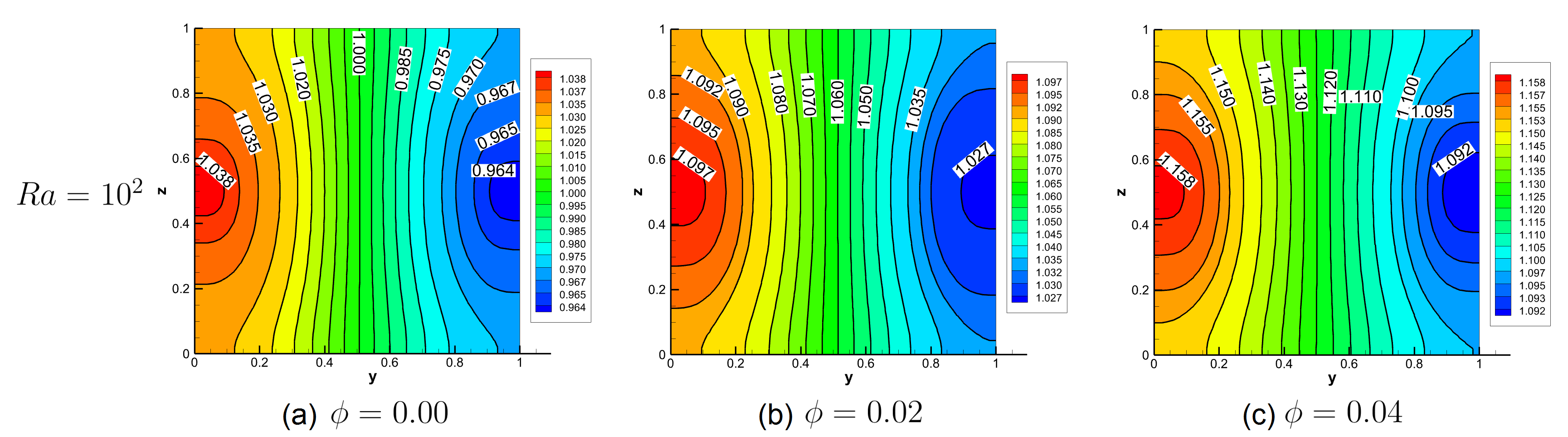}%
    \captionsetup{skip=2pt}%
  \end{subfigure}%   
  \hspace*{\fill}%          % empty line absolutely necessary!

  \vspace*{8pt}%
  \hspace*{\fill}%  
  \begin{subfigure}{1.0\textwidth}     % start subfigure 1
    %\belowcaptionskip=8pt
    \centering
    \includegraphics[width=\textwidth]{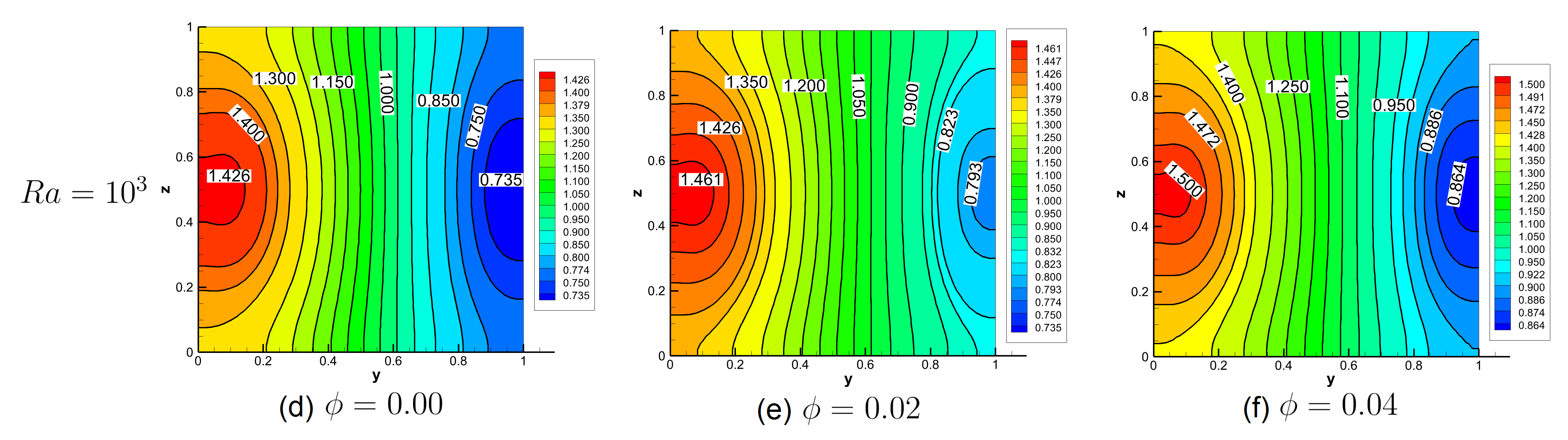}%
    \captionsetup{skip=2pt}%
  \end{subfigure}%   

  \vspace*{8pt}%
  \hspace*{\fill}%  
  \begin{subfigure}{1.0\textwidth}     % start subfigure 1
    %\belowcaptionskip=8pt
    \centering
    \includegraphics[width=\textwidth]{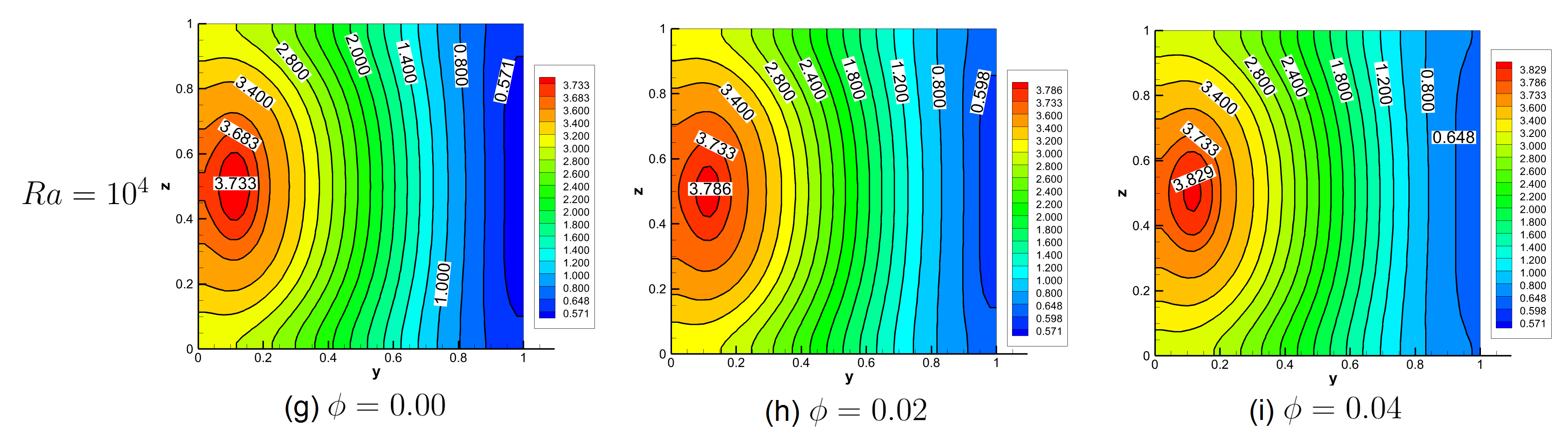}%
    \captionsetup{skip=2pt}%
  \end{subfigure}%   
  \hspace*{\fill}%          % empty line absolutely necessary!

  \vspace*{8pt}%
  \hspace*{\fill}%  
  \begin{subfigure}{1.0\textwidth}     % start subfigure 1
    %\belowcaptionskip=8pt
    \centering
    \includegraphics[width=\textwidth]{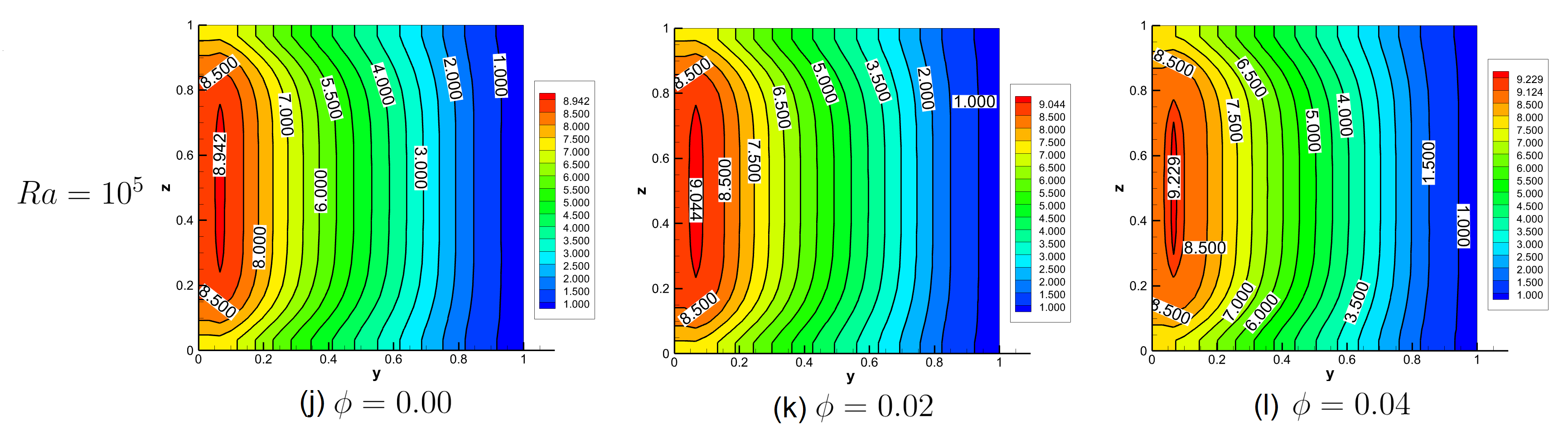}%
    \captionsetup{skip=2pt}%
  \end{subfigure}%   
  \hspace*{\fill}%          % empty line absolutely necessary!
  \vspace*{2pt}%
  \hspace*{\fill}%  
  \caption{Case 1: Local Nusselt number ($Nu_{\text{L}}$) on the heated wall ($x=0$): (a) $Ra=10^2$ (b) $Ra=10^3$ (c) $Ra=10^4$ (d) $Ra=10^5$}
  \label{fig:LC_NU_P1}
\end{figure}

Figure \ref{fig:LC_NU_P1} shows the distribution of the local Nusselt number for Case 1 on the heated wall at $x = 0.0$, across different Rayleigh numbers ($Ra$) and nanoparticle volume fractions ($\phi$). These variations provide valuable information about the system's convective heat transfer behavior. At a lower Rayleigh number ($Ra = 10^2$), the contours exhibit a symmetrical distribution with respect to the $y = 0.5$ plane. However, as $Ra$ increases,  a notable asymmetry emerges, particularly at $Ra = 10^5$, where the distribution of the Nusselt numbers clearly reflects the influence of strong convective flows. The highest values of the Nusselt number are observed near the $y = 0.0$ (bottom side of the cavity), and these maxima further increase with higher $\phi$, underscoring the enhanced heat transfer rates with increasing nanoparticle concentration. The approximate maximum values of $Nu_L$ (Figure \ref{fig:LC_NU_P1}) at $Ra = 10^5$ are 8.942, 9.044, and 9.229 for $\phi = 0.0$, 0.02, and 0.04, respectively. This trend illustrates the direct impact of $\phi$ on increasing heat transfer efficiency. Interestingly, the $Nu_L$ values near $y=0.0$ and $z = 0.5$ are consistently higher compared to those near the adiabatic walls, suggesting that the lateral adiabatic boundaries suppress convective heat transfer inside the cubic cavity. This influence is particularly pronounced as $\phi$ increases, leading to a significant rise in the local Nusselt number values. Specifically, we observe an approximate 11.5\%, 5.1\%, 2.5\%, and 3.2\% increase in the maximum local Nusselt number for $Ra = 10^2$, $10^3$, $10^4$, and $10^5$, respectively, as $\phi$ increases from 0.00 to 0.04.
These findings emphasize the complex interaction between $Ra$, $\phi$, and localized heat transfer inside the cavity for Case 1. 
%%%%%%%%%%%%%%%%%%%%%%%%%%%%%%%%%%%%%%%%%%%%%%%%%%%%%%%%%%%%%%%%%%%%%%%%%%%%%%%%%%%%%%%%%%%%%
%% Streamlines at 3 planes %%%
%%%%%%%%%%%%%%%%%%%%%%%%%%%%%%%%%%%%%%%%%%%%%%%%%%%%%
%%%%%%%%%%%%%%%%%%%%%%%%%%%%%%%%%%%%%%%%%%%%%%%%%%%%%%%%%%%%%%%%%%%%%%%%%%%%%%%

%%%%%%%%%%%%%%%%%%%%%%%%%%%%%%%%%%%%%%%%%%%%%%%%%%%%%%%%%%%%%%%%%%%%%%%%%%%%%%%
\begin{figure}[htbp]
 \centering
 \vspace*{0pt}%
 \hspace*{\fill}% 
\begin{subfigure}{1.0\textwidth}     % start subfigure 1
    %\belowcaptionskip=8pt
    \centering
    \includegraphics[width=\textwidth]{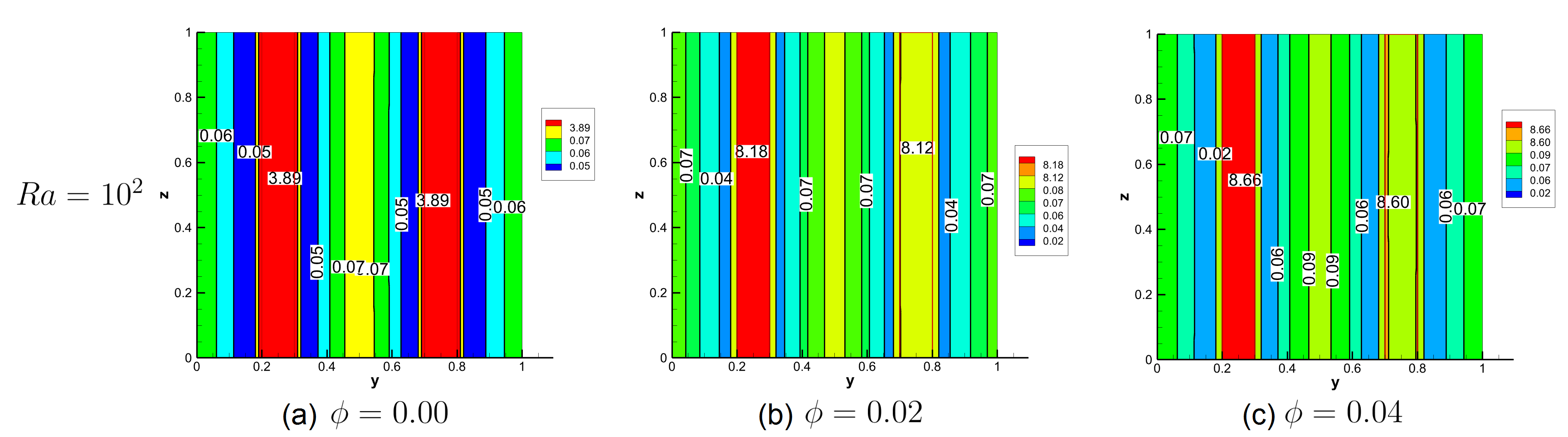}%
    \captionsetup{skip=2pt}%
  \end{subfigure}%   
  \hspace*{\fill}%          % empty line absolutely necessary!

  \vspace*{8pt}%
  \hspace*{\fill}%  
  \begin{subfigure}{1.0\textwidth}     % start subfigure 1
    %\belowcaptionskip=8pt
    \centering
    \includegraphics[width=\textwidth]{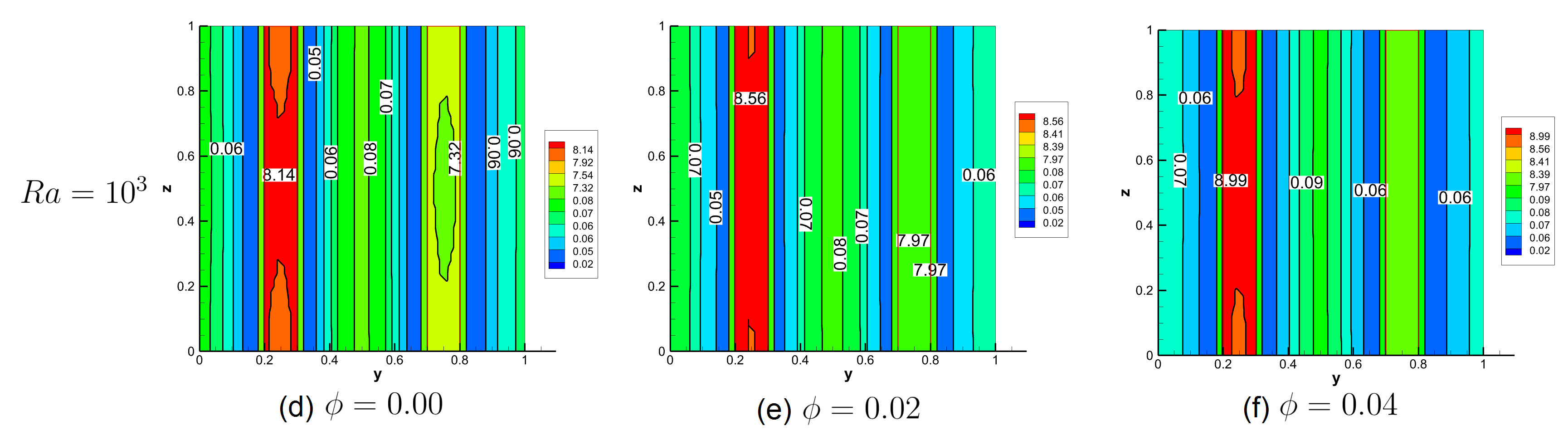}%
    \captionsetup{skip=2pt}%
  \end{subfigure}%   

  \vspace*{8pt}%
  \hspace*{\fill}%  
  \begin{subfigure}{1.0\textwidth}     % start subfigure 1
    %\belowcaptionskip=8pt
    \centering
    \includegraphics[width=\textwidth]{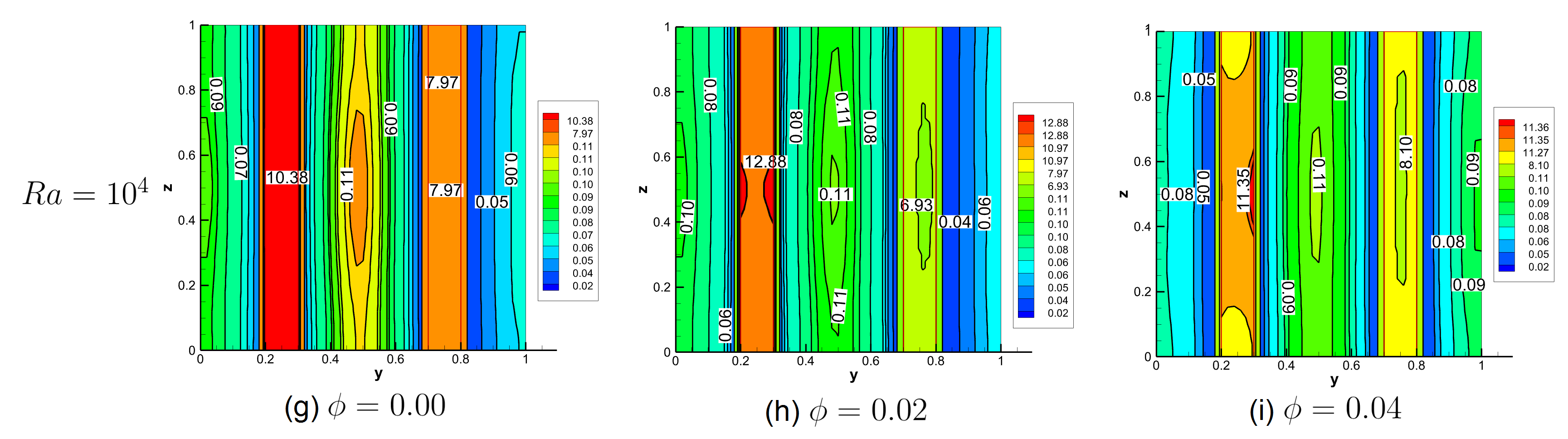}%
    \captionsetup{skip=2pt}%
  \end{subfigure}%   
  \hspace*{\fill}%          % empty line absolutely necessary!

  \vspace*{8pt}%
  \hspace*{\fill}%  
  \begin{subfigure}{1.0\textwidth}     % start subfigure 1
    %\belowcaptionskip=8pt
    \centering
    \includegraphics[width=\textwidth]{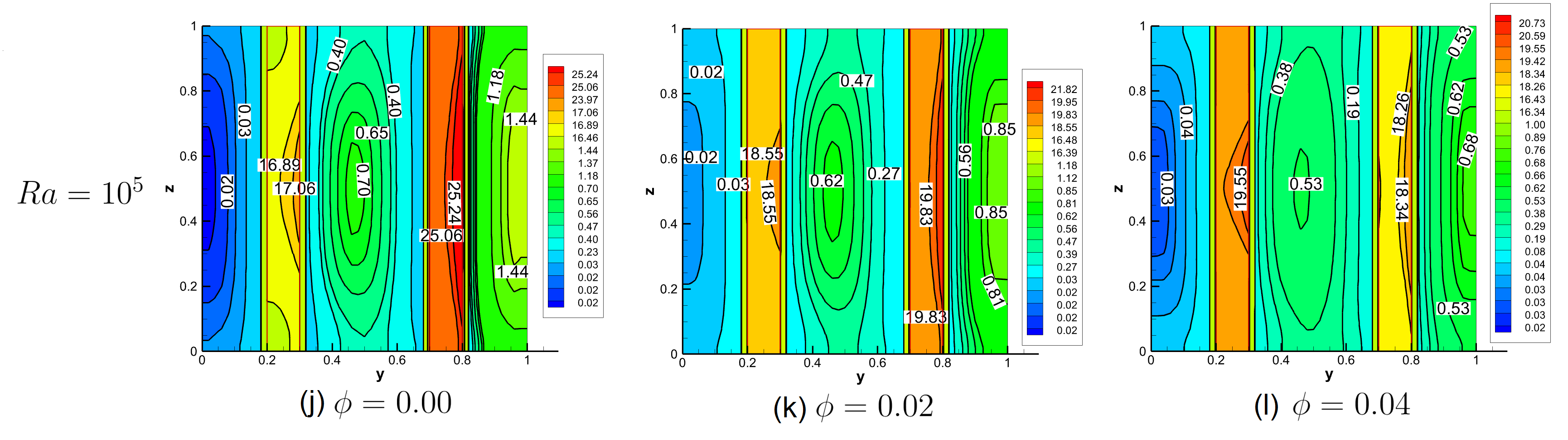}%
    \captionsetup{skip=2pt}%
  \end{subfigure}%   
  \hspace*{\fill}%          % empty line absolutely necessary!
  \vspace*{2pt}%
  \hspace*{\fill}%  
  \caption{Case 2: Local Nusselt number ($Nu_{\text{L}}$) on the heated wall ($x=0$): (a) $Ra=10^2$ (b) $Ra=10^3$ (c) $Ra=10^4$ (d) $Ra=10^5$}
  \label{fig:LC_NU_P2}
\end{figure}

Figure \ref{fig:LC_NU_P2} presents the local Nusselt number contours for Case 2. At $Ra=10^2$ and $10^3$, the contours appear almost as straight lines, reflecting a predominantly conductive heat transfer regime. As $Ra$ increases to $10^4$, the contours between the two conducting fins become slightly stretched into circular curves, indicating the emergence of stronger convective heat transfer. Notably, the highest local Nusselt number values are observed at the points where the conducting fins are attached to the heated wall, across all $Ra \leq 10^4$. At $\phi = 0.04$, the approximate maximum $Nu_L$ values observed in Figure \ref{fig:LC_NU_P2} are 8.66, 8.99, and 11.35 for $Ra = 10^2$, $10^3$, and $10^4$, respectively. Among the two fins, the bottom fin consistently exhibits the maximum Nusselt number, a clear indication of more intense heat transfer in that region. Additionally, the increase in $\phi$ causes a corresponding rise in the maximum local Nusselt number, demonstrating the positive effect of nanoparticle concentration on heat transfer efficiency. There is approximately a 122.6\%, 10.44\%, and 9.34\% increase in the maximum $Nu_L$ at $Ra = 10^4$, $Ra = 10^3$, and $Ra = 10^2$, respectively, when transitioning from $\phi = 0.0$ to $\phi = 0.04$. At $Ra=10^5$, a significant shift occurs: the maximum local Nusselt number now appears near the base of the top fin, marking a reversal in the trend observed at lower $Ra$ values. This shift underscores the complex interaction between buoyancy forces and the thermal field, confirming the previously discussed isotherm patterns. The observed changes in the local Nusselt number contours at high $Ra$ further emphasize the critical role of the conducting fins in redistributing heat and enhancing convective heat transfer within the cavity.

We have also calculated the total Nusselt number ($Nu_{\text{T}}$) for both cases, with the values presented in Table \ref{Total_Nusselt_Number_Comparison}. The table includes the percentage change in $Nu_{\text{T}}$ for Case 2 relative to Case 1, providing a clear quantification of the heat transfer enhancement due to the addition of conducting fins at various $Ra$ and $\phi$ values. Notably, the inclusion of conducting fins can lead to an enhancement in heat transfer by up to 88.9\% ($Ra=10^2$, $\phi=0.0$). 
The enhancement percentage of $Nu_{T}$ in Case 2 decreases with increasing $Ra$ values for both pure fluid and nanofluid. This occurs because, at lower $Ra$, where conduction dominates, the conducting fins significantly boost heat transfer by enhancing conduction. However, it is important to observe that while the presence of conducting fins generally enhances $Nu_{\text{T}}$, this is not always the case. For instance, at $Ra=10^5$ and $\phi=0.04$, the addition of fins does not result in an increased heat transfer rate. This indicates that the combination of fins with nanofluid does not universally guarantee an enhancement in heat transfer efficiency.

\begin{figure}[htbp]
 \centering
 \vspace*{5pt}%
 \hspace*{\fill}% 
\begin{subfigure}{0.50\textwidth}     % start subfigure 1
    %\belowcaptionskip=8pt
    \centering
    \includegraphics[width=\textwidth]{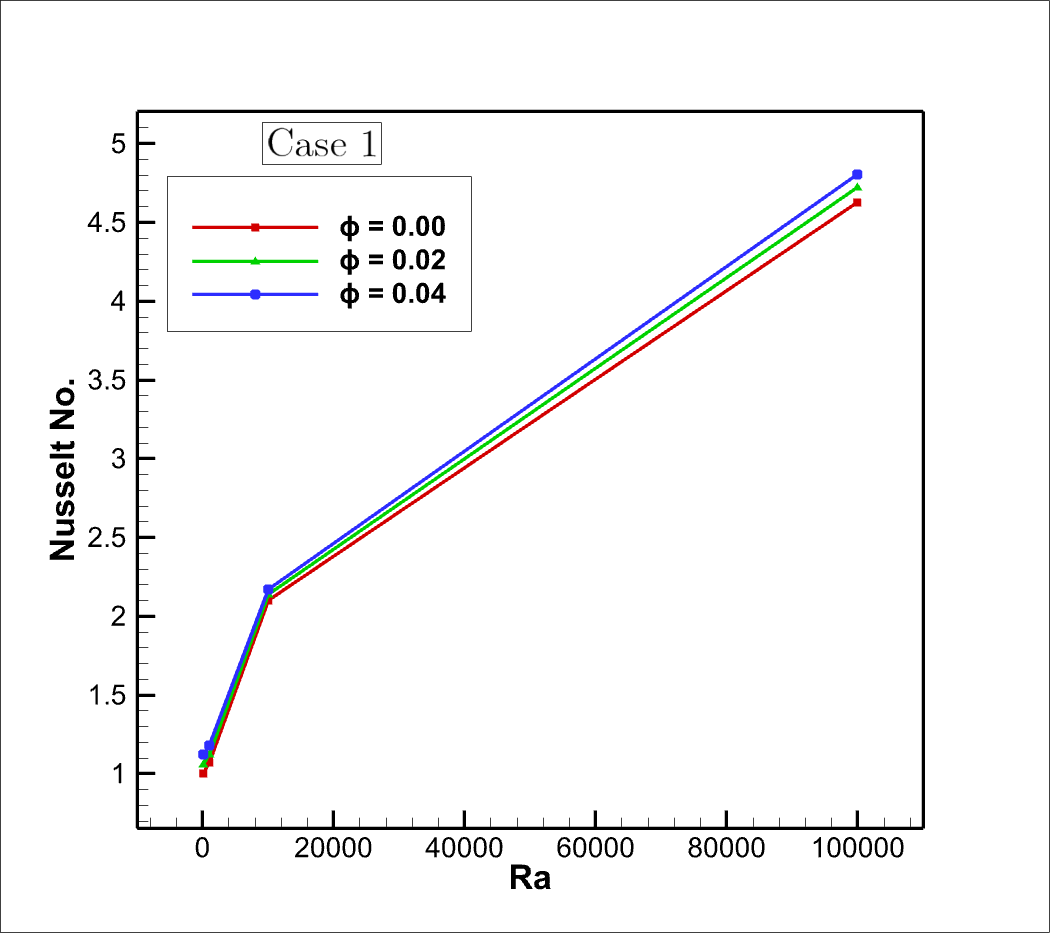}%
    \captionsetup{skip=5pt}%
    \caption{(a)}
    \label{fig:Ra_10^2_avg}
  \end{subfigure}%   
         % end subfigure 1
         % empty line absolutely necessary!
 \begin{subfigure}{0.50\textwidth}        % start subfigure 2
    %\belowcaptionskip=8pt
   \centering
    \includegraphics[width=\textwidth]{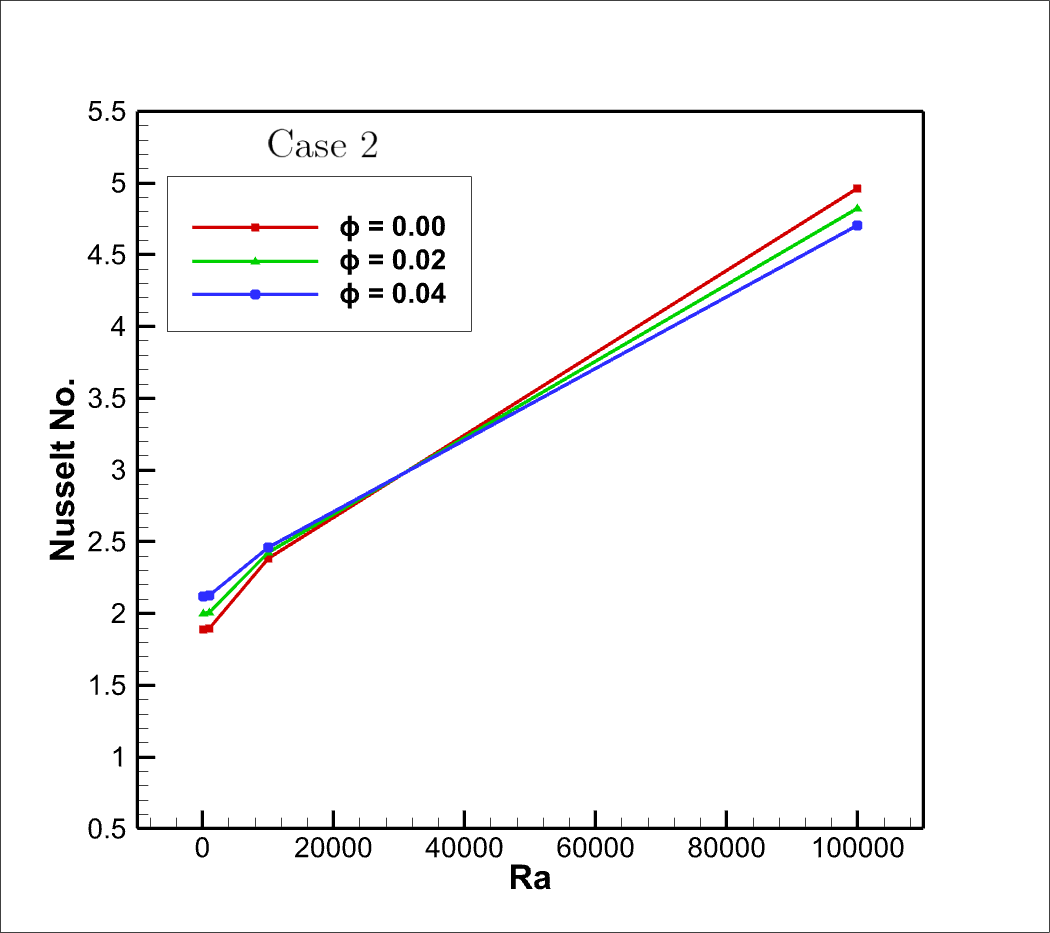}%
    \captionsetup{skip=5pt}%
    \caption{(b)}
    \label{fig:Ra_10^3_avg}
  \end{subfigure}%          % end subfigure 2  

   \caption{Variation of overall/total Nusselt number ($Nu_{\text{T}})$: (a) Case 1 (b) Case 2 }
  \label{fig:total_Nusselt_Number_1}
\end{figure}

To better visualize the impact of $Ra$ and $\phi$ on the total Nusselt number for both cases, we have also provided a graphical representation in Figure \ref{fig:total_Nusselt_Number_1}. From this figure, it is evident that as $Ra$ increases, the $Nu_{\text{T}}$ values rise consistently for both cases. However, the effect of $\phi$ varies between the cases. In Case 1, $Nu_{\text{T}}$ consistently increases with $\phi$ at every specific $Ra$ value. In Case 2, this trend holds for lower $Ra$ values ($\leq 10^4$) but reverses at $Ra=10^5$. This reversal suggests that the addition of nanoparticles does not always enhance heat transfer rates, particularly in combination with conducting fins at high $Ra$.
%%%%%%%%%%%%%%%%%%%%%%%%%%%%%%%%%%%%%%%%%%%%%%%%%%%%%%%%%%%%%%%%%%%%%%%%%%%%%%%%%%%%%%%%%%%%%

{\tiny \begin{table}[htbp] \footnotesize
\caption{\small Total Nusselt Number $Nu_{\text{T}}$ computed on the heated wall on $51\times51\times51$ grid size for Case 1 (C1) and Case 2 (C2).}\label{Total_Nusselt_Number_Comparison}
\centering
 \begin{tabular}{cccccccccccc}  \hline \hline
 &  &   \multicolumn{3}{c}{$\phi = 0.00$}    &\multicolumn{3}{c}{$\phi = 0.02$}    & \multicolumn{3}{c}{$\phi = 0.04$}  \\ \hline 
& $Ra$ &   C1  & C2  & change (\%)    & C1   &  C2 &   change (\%) & C1   &  C2 &   change (\%)\\ \hline 
& $ 10^2$  & 1.000 &  1.889    & +88.9 & 1.061   & 2.003 & +88.7 & 1.125& 2.121 & +88.5 \\
& $ 10^3$   & 1.074 & 1.899    & +76.8  & 1.125  & 2.010 & +78.6 & 1.179 & 2.127 & +80.4 \\
& $ 10^4$   & 2.102  & 2.386  & +13.5 & 2.137 & 2.425 & +13.4 & 2.170 & 2.462 & +13.4\\
& $ 10^5$   &  4.627 & 4.965   &  +7.3 & 4.719  & 4.825 & +2.24 & 4.807 & 4.705 & -2.1\\

\hline\hline
 \end{tabular}
\end{table}
}
\newpage
\section{Conclusion}
\label{sec:Conclusion}
This study marks the first application of the Higher-Order Super-Compact (HOSC) scheme to investigate the complex phenomena of natural convection within a 3D cavity incorporating conducting fins and nanofluid. The comparison with existing 3D results, both quantitatively and qualitatively, demonstrates the scheme's effectiveness in accurately capturing complex flow phenomena, providing strong motivation for further investigation. Our analysis reveals several key insights. Firstly, the addition of conducting fins significantly alters the flow patterns and heat transfer rates, leading to the formation of multiple secondary vortices and the distortion of the primary vortex structure. Secondly, the study underscores that the integration of nanoparticles does not unconditionally enhance heat transfer. While higher nanoparticle concentration generally boosts the local and total Nusselt numbers, the effectiveness of this enhancement is highly dependent on the Rayleigh number and the specific configuration of the system. Notably, the introduction of conducting fins combined with nanofluid at high Rayleigh numbers ($Ra=10^5$) can lead to a counterintuitive decrease in heat transfer efficiency, challenging the conventional understanding of nanofluid behavior. These findings not only advance our understanding of heat transfer in complex systems but also validate the potential of the HOSC scheme as a powerful tool for simulating such phenomena with high accuracy. This study opens new avenues for optimizing the design of heat transfer systems, particularly in applications involving advanced cooling technologies and thermal management. The novel insights gained here underscore the importance of considering a multitude of factors—such as Rayleigh number, nanoparticle concentration, and conducting elements—when aiming to enhance heat transfer performance.

In conclusion, the effective application of the new HOSC scheme in this research provides a robust foundation for future studies, paving the way for more comprehensive explorations of 3D fluid flow and heat transfer problems in complex geometries and with advanced materials like nanofluids.
\vspace{11pt}\\
{\textbf{Author Declaration}}\\
The authors have no conflicts to disclose.
\vspace{11pt}\\
{\textbf{Data Availability}}\\
The data that support the findings of this study are available from the corresponding author upon reasonable request.
\vspace{11pt}\\
{\textbf{Funding}}\\
This work is supported by the \textbf{Scheme for Promotion of Academic and Research Collaboration (SPARC)} under the Ref No. \textbf{SPARC/2024-2025/CAMS/P2742 Dated 02/05/2024} .




% \end{thebibliography}

\end{document}